\documentclass[11pt]{article}
\usepackage[utf8]{inputenc}
\usepackage{amsmath,amsfonts,amssymb}
\usepackage{geometry}
\usepackage{booktabs}
\usepackage{array}
\usepackage{mathtools}
\usepackage{bm}
\usepackage{xcolor}
\usepackage{enumitem}
\usepackage{algorithm}
\usepackage{algorithmic}
\usepackage{setspace}
\usepackage{tikz}
\usepackage{graphicx}
\usepackage{caption}
\usepackage{subcaption}
\usepackage{comment}

\usepackage{times}

\newcommand{\bu}{\mathbf{u}}                   % State vector
\newcommand{\blf}{\mathbf{f}}                  % Flux in x-direction
\newcommand{\bg}{\mathbf{g}}                   % Flux in y-direction
\newcommand{\bs}{\mathbf{s}}                   % Source term

\newcommand{\bmu}{\boldsymbol{\mu}}            % Parameter vector
\newcommand{\hL}{h_{\!L}}                      % Upstream depth
\newcommand{\hR}{h_{\!R}}                      % Downstream depth

\newcommand{\Nx}{N_x}                          % Spatial points in x
\newcommand{\Ny}{N_y}                          % Spatial points in y
\newcommand{\dx}{\Delta x}                     % Grid spacing in x
\newcommand{\dy}{\Delta y}                     % Grid spacing in y
\newcommand{\dt}{\Delta t}                     % Time step

\newcommand{\bV}{\mathbf{V}}                   % POD/ROM basis
\newcommand{\bU}{\mathbf{U}}                   % Left singular vectors
\newcommand{\bW}{\mathbf{W}}                   % Spatial factor (Tucker)
\newcommand{\Sig}{\boldsymbol{\Sigma}}         % Singular values

\newcommand{\bX}{\mathbf{X}}                   % Snapshot matrix
\newcommand{\Ccore}{\mathcal{C}}               % Core tensor (Tucker)
\newcommand{\QQ}{\mathcal{Q}}                  % Snapshot tensor

\newcommand{\epsloc}{\epsilon_{\text{loc}}}

\newcommand{\Ntau}{N_{\tau_c}}                 % Samples for yield stress
\newcommand{\NL}{N_L}                          % Samples for h_L
\newcommand{\NR}{N_R}                          % Samples for h_R
\newcommand{\Ntheta}{N_{\theta}}               % Samples for slope

\newcommand{\OmegaD}{\Omega}

\newcommand{\pd}[2]{\frac{\partial #1}{\partial #2}}

\newcommand{\tauc}{\tau_c}                     % Yield stress
\newcommand{\thet}{\theta}                     % Bed slope angle

\newcommand{\itc}[1]{{\color{blue}#1}}

\newtheorem{remark}{Remark}

\begin{document}

% ==============================
% TITLE AND AUTHORS
% ==============================

\title{Reduced-Order Modeling of Parameterized Visco-Plastic Shallow Flows}

\author{
Md Rezwan Bin Mizan \thanks{Dept. of Mathematics, University of Houston, Houston, TX 77204,
mbinmiza@cougarnet.uh.edu} \quad
Ilya Timofeyev\thanks{Dept. of Mathematics, University of Houston, Houston, TX 77204, itimofey@cougarnet.uh.edu}
 \quad
Maxim Olshanskii \thanks{Dept. of Mathematics, University of Houston, Houston, TX 77204, maolshanskiy@uh.edu}
}

\date{}

\maketitle

% ==============================
% ABSTRACT
% ==============================

\begin{abstract}
We propose a non-intrusive reduced-order modeling framework for parametrized visco-plastic free-surface flows governed by a shallow-water formulation of Herschel--Bulkley fluids. These flows exhibit strong nonlinearities, non-smooth rheology, moving fronts, and yield surfaces, making efficient surrogate modeling particularly challenging. 
To address this challenge, we employ a tensor-based approach in which the solution manifold is approximated using a low-rank representation obtained via higher-order singular value decomposition of snapshot data over a structured parameter space.

The resulting tensorial reduced-order model (TROM) enables rapid online evaluation by directly reconstructing solution trajectories from the compressed representation, thereby avoiding the need to perform time integration of a reduced dynamical system. The proposed non-intrusive framework can be interpreted as an encoder--decoder architecture with a compressed latent representation and efficient multilinear decoding. Numerical experiments demonstrate that the proposed approach accurately captures key flow features, including front propagation, plug and shear regions, and near-stopping dynamics, while achieving substantial computational speedups relative to full-order simulations.
\end{abstract}


%%%%%%%%%%%%%%%%%%%%%%%%%%%%%%%%%%%
% Section 1
% INTRODUCTION
%%%%%%%%%%%%%%%%%%%%%%%%%%%%%%%%%%%
\section{Introduction}

Reduced order modeling (ROM) has become an essential tool in
simulations of complex fluid flows, particularly in applications that require repeated evaluations of high-fidelity models. Such situations arise naturally in inverse problems, parameter estimation, uncertainty quantification, and optimization, where the governing equations must be solved many times for different parameter values. For non-Newtonian fluids, whose rheology introduces strong nonlinearities and often non-smooth behavior, {repeated simulations of a full-order model} can be prohibitively expensive, making efficient surrogate models especially valuable.

In this work, we focus on ROM techniques for visco-plastic free-surface flows. These flows arise in a variety of natural and industrial processes, including debris flows, mudslides, lava flows, and the transport of slurries and pastes. Accurate and rapid prediction of key features such as flow extent, stopping time, and the formation of plug and yield regions is 
%critical 
{crucial} for hazard assessment and process control. In particular, fast predictive capabilities are highly desirable in the context of emerging digital twin (DT) technologies for geophysical and environmental systems \cite{ugliotti2023landscape,hazeleger2024digital}, where real-time or near-real-time simulations are integrated with observational data to support decision-making. While digital twins have been actively developed for Newtonian fluid systems \cite{yang2024data}, their extension to non-Newtonian and, in particular, visco-plastic flows remains largely unexplored.
{Here we investigate a key component of the digital-twin paradigm: a parametric reduced-order model for fast evaluation.}

The development of ROMs for Newtonian fluid models has seen significant progress over the past \itc{few} decades, with numerous approaches based on Proper Orthogonal Decomposition (POD), reduced basis methods, and operator inference (see, e.g., \cite{quarteroni2015reduced, benner2015survey, ballarin2016supremizer, rowley2005model}). In contrast, the literature on ROMs for non-Newtonian flows is considerably more limited. Existing works have primarily focused on specific non-Newtonian rheologies, including viscoelastic and generalized Newtonian models, \cite{wang2020pod, chetry2023stabilization, girault2023thermorheological, oishi2024nonlinear}, while ROMs for visco-plastic fluids remain scarce. The non-smoothness introduced by yield stress, along with the presence of unyielded (plug) regions, poses additional challenges that are not encountered in Newtonian settings.

In this paper, we address the problem of constructing efficient and accurate ROMs for parametrized visco-plastic shallow free-surface flows. Our focus is on flows governed by a two-dimensional shallow flow model for Herschel--Bulkley fluids \cite{muchiri2025derivation}.

Depth-averaged models for visco-plastic free-surface flows are typically derived using either lubrication-type approximations or shallow water (Saint-Venant) approaches. Lubrication models, developed under the assumption of slow flows with negligible inertia, lead to reduced advection–diffusion equations and have been widely used; see, e.g., \cite{huang1998herschel,balmforth1999consistent, balmforth2006viscoplastic, ancey2009dam}. In contrast, shallow water equation (SWE) models are obtained by depth-integration of the full mass and momentum balances and retain inertial effects; see, e.g., \cite{laigle1997numerical,fernandez2010shallow,fernandez2023multilayer,muchiri2025derivation}.

Recent comparative studies~\cite{muchiri2025comparing} show that while both approaches perform similarly in low-inertia regimes, lubrication models tend to overpredict front propagation and fail to capture transient dynamics when inertial effects become significant, whereas SWE models provide more accurate predictions, particularly during rapidly evolving phases such as dam-break flows and in the presence of more complex topography. In this work, we adopt the model introduced in \cite{muchiri2025derivation,muchiri2024numerical}, which provides an SWE formulation for Herschel--Bulkley fluids with improved treatment of gravity and basal friction terms. This model has been shown to achieve good agreement with experimental data across a range of flow regimes~\cite{muchiri2025comparing} while remaining computationally tractable, making it a natural choice as the high-fidelity FOM.

Constructing ROMs for such systems presents several challenges. First, the governing equations are hyperbolic, which leads to the presence of sharp fronts. Second, the nonlinear dependence of the fluxes and source terms on the solution and parameters is particularly strong due to the Herschel--Bulkley rheology. Third, accurately capturing delicate flow features such as the evolution of yield surfaces, the extent of plug regions, and the final stopping time is non-trivial, especially when these features depend sensitively on multiple parameters, including yield stress, initial conditions, and bed slope. These aspects make standard linear reduction techniques insufficient without careful adaptation.

To address these challenges, we follow the approach of~\cite{mamonov2022interpolatory} and propose a non-intrusive reduced-order model (ROM) that employs a low-rank tensor decomposition (LRTD) as the model reduction technique. Specifically, we apply the higher order singular value decomposition (HOSVD) to construct a compact representation of the high-dimensional solution manifold arising from parametric variations. The proposed tensorial reduced order modeling (TROM) framework operates in two stages. In the offline phase, we generate solution snapshots over a structured parameter grid and compute the HOSVD to extract dominant modes in space, time, and parameter dimensions. In the online phase, the reduced representation enables rapid evaluation of the solution for new parameter values.

While the work in \cite{mamonov2022interpolatory, mamonov2024tensorial} considered an \emph{intrusive} approach based on Galerkin projection, where the governing equations are projected onto the reduced space and integrated in time, the TROM introduced here features a \emph{non-intrusive} online reconstruction that directly evaluates the reduced tensor representation to recover full solution trajectories without solving a lower-dimensional dynamical system. This non-intrusive TROM can be interpreted as an encoder–decoder scheme, with an offline “training” phase that provides a nonlinear encoder and a multilinear decoder.
Within this interpretation, the collection of snapshots and the computation of a low-rank tensor decomposition (LRTD) can be viewed as training, while for any new parameter value, interpolation in the tensor low-rank format followed by a truncated SVD of the resulting core matrix constitutes the encoding step, which is then followed by decoding to recover the corresponding space–time solution; cf. Section~\ref{sec:3.2}.

The remainder of the paper is organized as follows. In Section~\ref{sec:setup}, we present the mathematical formulation of the visco-plastic shallow flow model and describe the full-order numerical scheme. Section~\ref{sec:trom} introduces the tensor-based ROM framework, including the construction of snapshot tensors and the Tucker decomposition. Numerical experiments demonstrating the accuracy and efficiency of the proposed approach are presented in Section~\ref{sec:num}. Finally, conclusions and perspectives for future work are discussed in Section~\ref{sec:conc}.

%%%%%%%%%%%%%%%%%%%%%%%%%%%%%%%%%%%
% Section 2
% MATHEMATICAL MODEL AND FULL-ORDER MODEL
%%%%%%%%%%%%%%%%%%%%%%%%%%%%%%%%%%%
\section{Problem setup  and FOM}
\label{sec:setup}

\subsection{Two-Dimensional Herschel--Bulkley Shallow Water Equations}

As a full-order model, we consider the shallow water equations for visco-plastic fluids introduced in~\cite{muchiri2024numerical}. This model provides a computationally efficient framework for describing free-surface visco-plastic flows while retaining the key rheological features of Herschel--Bulkley fluids. In particular, it captures the interplay between gravity-driven motion and yield-stress effects through depth-averaged variables, making it suitable for large-scale geophysical applications such as debris flows and mudflows. 
The model incorporates $\Lambda$-corrective momentum fluxes and multi-regime gravity formulations to accurately capture the complex rheological behavior of Herschel--Bulkley fluids.

%\subsubsection{Governing Equations}

The governing equations are given by the two-dimensional shallow water system written in conservative form:
\begin{equation}
\frac{\partial \bu}{\partial t} + \frac{\partial \blf(\bu)}{\partial x} + \frac{\partial \bg(\bu)}{\partial y} = \bs_g + \bs_f,
\label{eq:2d_hb_conservative}
\end{equation}
with the state vector
\begin{equation}
\bu = \begin{bmatrix} h \quad q_x \quad q_y \end{bmatrix}^T.
\label{eq:state_vector_2d}
\end{equation}
The source terms $\bs_g$ and $\bs_f$ accounts for gravity and friction. They are defined later.

The flux vectors in the $x$ and $y$ directions depend on the flow depth $h = h(\mathbf{x}, t)$ and the fluid discharges $q_x = q_x(\mathbf{x}, t)$ and $q_y = q_y(\mathbf{x}, t)$:
\begin{equation}
\blf(\bu) = \begin{bmatrix} 
q_x \\[4pt]
\dfrac{q_x^2}{h} + \dfrac{1}{2} g h^2 \cos\theta + \Lambda_{xx} \\[6pt]
\dfrac{q_x q_y}{h} + \Lambda_{xy}
\end{bmatrix}, \quad 
\bg(\bu) = \begin{bmatrix} 
q_y \\[4pt]
\dfrac{q_x q_y}{h} + \Lambda_{xy} \\[6pt]
\dfrac{q_y^2}{h} + \dfrac{1}{2} g h^2 \cos\theta + \Lambda_{yy}
\end{bmatrix}.
\label{eq:flux_vectors_2d}
\end{equation}

The discharges in the $x$ and $y$ directions are defined as $q_x = h u$ and $q_y = h v$, where $u$ and $v$ denote the depth-averaged velocity components. The terms $\Lambda_{xx}$, $\Lambda_{yy}$, and $\Lambda_{xy}$ in the definitions of $\blf$ and $\bg$ represent corrective momentum fluxes, which will be defined later.

\paragraph{Herschel--Bulkley Rheology.}

The Herschel--Bulkley constitutive law describes a visco--plastic fluid characterized by three parameters: the yield stress $\tauc \geq 0$, the consistency index $K > 0$, and the flow index $n > 0$. The relationship between the shear stress $\tau$ and the shear rate $\dot{\gamma}$ is given by
\begin{equation}
\tau = \tauc + K\dot{\gamma}^n.
\label{eq:hb_constitutive}
\end{equation}

\begin{figure}[h]
\centering
\includegraphics[width=0.8\textwidth]{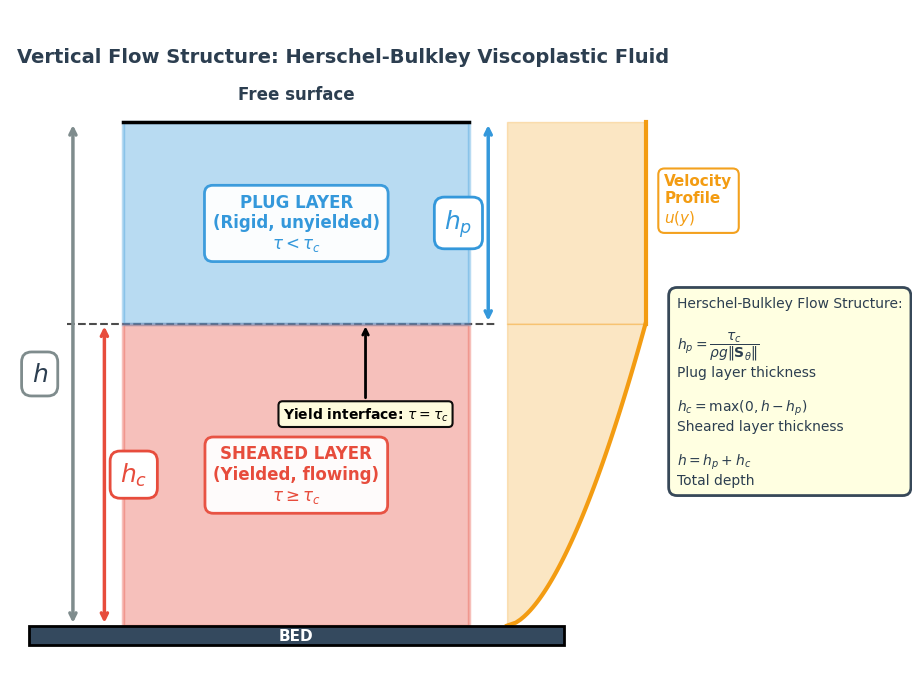}
\caption{Vertical flow structure for a Herschel--Bulkley visco--plastic fluid, showing the plug layer (rigid, unyielding region where $\tau < \tau_c$), the sheared layer (yielded, flowing region where $\tau \geq \tau_c$), and the characteristic velocity profile, with constant velocity in the plug and a parabolic profile in the sheared layer.}
\label{fig:flow_structure}
\end{figure}

The defining characteristic of yield-stress fluids is the existence of a critical stress threshold below which the material behaves as a rigid solid. In free-surface shallow flows, this leads to a vertical flow structure consisting of an upper rigid plug layer of thickness $h_p(\mathbf{x},t)$ and a lower sheared layer of thickness $h_c(\mathbf{x},t)$ (see Figure~\ref{fig:flow_structure}). Following~\cite{muchiri2024numerical}, these are defined as
\begin{equation}
h_p = \frac{\tauc}{\rho g \|\mathbf{S}_\theta\|}, \quad 
h_c = \max(0, h - h_p),
\label{eq:plug_shear_thickness}
\end{equation}
where $\mathbf{S}_\theta$ is the driving slope vector defined below and $\rho$ is the fluid density.

\paragraph{Corrective Momentum Fluxes.}

The terms $\Lambda_{xx}$, $\Lambda_{yy}$, and $\Lambda_{xy}$ in equation~\eqref{eq:flux_vectors_2d} represent corrective momentum fluxes that account for the non-uniform velocity profile. These terms are given by
\begin{equation}
\Lambda_{xx} = C_m \Lambda_x^{2m} h^{2m+3}, \quad 
\Lambda_{yy} = C_m \Lambda_y^{2m} h^{2m+3}, \quad 
\Lambda_{xy} = C_m \Lambda_x^m \Lambda_y^m h^{2m+3},
\label{eq:lambda_corrections}
\end{equation}
where
\begin{equation}
C_m = \frac{1}{(2m+3)(m+2)^2}, \quad m = \frac{1}{n},
\label{eq:cm_parameter}
\end{equation}
and $n$ denotes the power-law index. The parameters $\Lambda_x$ and $\Lambda_y$ are defined as
\begin{equation}
\Lambda_x = \frac{K}{\rho g |S_{\theta,x}|}, \quad 
\Lambda_y = \frac{K}{\rho g |S_{\theta,y}|},
\label{eq:lambda_parameters}
\end{equation}
where $K$ is the consistency index, $g$ is the gravitational acceleration, and $S_{\theta,x}$ and $S_{\theta,y}$ denote the components of the driving slope vector.

\paragraph{Gravity Source Term.}

The gravity source term $\bs_g$ accounts for gravitational effects. For a flat bed ($b = 0$), as considered here, \cite{muchiri2024numerical} gives
\begin{equation}
\bs_g = \begin{bmatrix} 
0 \\[4pt]
- g h \sec\thet \dfrac{\partial h}{\partial x} \\[6pt]
- g h \sec\thet \dfrac{\partial h}{\partial y}
\end{bmatrix},
\label{eq:gravity_source}
\end{equation}
where $\thet$ is the mean slope angle and $\sec\thet = 1/\cos\thet$.

The driving slope vector for a flat bed is given by
\begin{equation}
\mathbf{S}_\theta = \begin{bmatrix} S_{\theta,x} \\ S_{\theta,y} \end{bmatrix} 
= \begin{bmatrix} 
\sin\thet - \cos\thet \dfrac{\partial h}{\partial x} \\[4pt]
- \cos\thet \dfrac{\partial h}{\partial y}
\end{bmatrix}.
\label{eq:driving_slope}
\end{equation}
In \eqref{eq:plug_shear_thickness}, we define $\|\mathbf{S}_\theta\| = \sqrt{S_{\theta,x}^2 + S_{\theta,y}^2}$.

\paragraph{Friction Source Term.}
The friction source term $\bs_f$ represents the basal shear stress arising from the Herschel--Bulkley rheology. The basal shear stress corresponds to the tangential force exerted by the fluid on the underlying surface due to viscous effects. This term governs energy dissipation and controls the flow resistance, thereby influencing the velocity, stopping behavior, and overall dynamics of the flow. {The friction term is given by}
\begin{equation}
\bs_f = -\frac{1}{\rho}\begin{bmatrix} 0 \quad \tau_{b,x} \quad \tau_{b,y} \end{bmatrix}^T,
\label{eq:friction_source}
\end{equation}
where $\tau_{b,x}$ and $\tau_{b,y}$ denote the components of the basal shear stress vector.
The basal shear stress components are expressed as~\cite{muchiri2024numerical}
\begin{equation}
\tau_{b,k} = K \left(\frac{K}{\rho g |S_{\theta,k}|}\right)^{m-1} \frac{h_p + h_c}{D(h)} q_k, \quad k \in \{x,y\},
\label{eq:basal_shear_stress}
\end{equation}
where the denominator function is
\begin{equation}
D(h) = h_c^{m+1} \left(\frac{h}{m+1} - \frac{h_c}{(m+1)(m+2)}\right) + C h^{m+1}.
\label{eq:denominator_function}
\end{equation}
Here, $C$ is the basal slip coefficient. In this study, we set $C = 0$, corresponding to a no-slip condition at the bed.

\subsection{Problem setup}

\begin{figure}[h]
\centering
\includegraphics[width=0.9\textwidth]{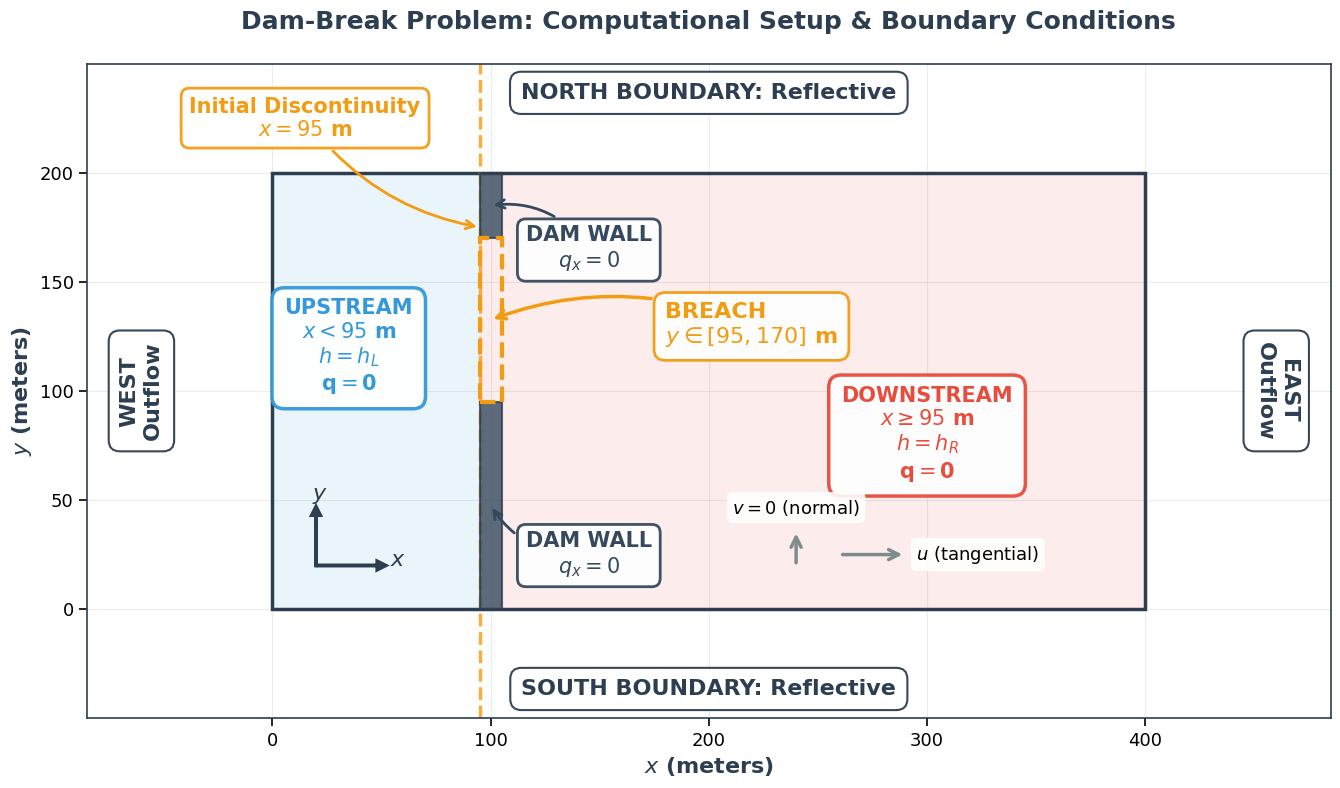}
\caption{Computational domain and boundary conditions for the two-dimensional visco-plastic dam-break problem. The domain shows the initial flow depth discontinuity at $x = 95$ m, the breach opening in the dam wall, and the boundary conditions applied at all domain boundaries.}
\label{fig:dam_setup}
\end{figure}

The computational domain $\OmegaD = [0, L_x] \times [0, L_y]$ represents a rectangular channel with dimensions $L_x \times L_y = 400 \times 200$ m. The dam is located at $x_d = 100$\,m and has a thickness of $\Delta x_d = 10$\,m. 
The breach opening is a portion of the dam defined by $y \in [y_0, y_1] = [95, 170]$\,m and extends over $x \in [x_d - \Delta x_d/2, x_d + \Delta x_d/2] = [95, 105]$\,m. See Figure~\ref{fig:dam_setup}.

\paragraph{Boundary Conditions.}
Following the setup in~\cite{muchiri2024numerical}, we consider free-slip and no-normal-flow boundary conditions at the top and bottom boundaries ($y = 0, L_y$):
\begin{equation}
\pd{h}{y} = 0, \quad \pd{q_x}{y} = 0, \quad q_y = 0,
\label{eq:bc_ns}
\end{equation}
outflow conditions at the left and right boundaries ($x = 0, L_x$):
\begin{equation}
\pd{h}{x} = 0, \quad \pd{q_x}{x} = 0, \quad \pd{q_y}{x} = 0,
\label{eq:bc_we}
\end{equation}
and zero normal flux through the intact portions of the dam:
\begin{equation}
q_x = 0 \quad \text{for } x \in [x_d - \Delta x_d/2, x_d + \Delta x_d/2], \, y \notin [y_0, y_1].
\label{eq:bc_gate}
\end{equation}

\paragraph{Initial Conditions.}
The dam-break problem is initialized with a discontinuous flow depth profile:
\begin{equation}
h(\mathbf{x}, 0) =
\begin{cases}
\hL & \text{if } x < x_d - \Delta x_d/2 \quad \text{(upstream)}, \\
\hR & \text{if } x > x_d + \Delta x_d/2 \quad \text{(downstream)},
\end{cases}
\label{eq:initial_h}
\end{equation}
and zero discharge:
\begin{equation}
q_x(\mathbf{x}, 0) = q_y(\mathbf{x}, 0) = 0 \quad \forall \mathbf{x} \in \OmegaD.
\label{eq:initial_q}
\end{equation}
\smallskip

\paragraph{Parameter space.}
We allow the following parameters to vary: the yield stress $\tauc$, the initial upstream and downstream depths $\hL,\, \hR$, and the mean slope angle $\thet$. These variations correspond to changes in the fluid rheology, the initial state, and the forcing terms in the equations. All of these factors substantially influence the critical flow statistics.

More specifically, we consider the following parameter ranges:
\begin{equation}\label{eq:paramDomain}
\tauc \in [0, 6000]\, \text{Pa}, \qquad
\hL \in [10, 15]\,\text{m}, \qquad
\hR \in [0, 2]\,\text{m}, \qquad
\thet \in [0^\circ, 4^\circ], 
\end{equation}
which define the parameter domain $\mathcal{P}$.

The parameter intervals in \eqref{eq:paramDomain} are discretized using a tensor-product grid. Specifically, we use
\[
\Ntau = 11, \quad \NL = 5, \quad \NR = 4, \quad \Ntheta = 5,
\]
sample points in the $\tauc$, $\hL$, $\hR$, and $\thet$ directions, respectively. 
The resulting parameter mesh consists of
\[
\Ntau \times \NL \times \NR \times \Ntheta = 11 \times 5 \times 4 \times 5 = 1100
\]
training parameter combinations. In this work, a uniform grid is employed in each parameter direction, with equally spaced points over the corresponding parameter intervals. 
% tau = 0, 600, 1200, 1800, 2400, 3000, 3600, 4200, 4800, 5400, 6000
% h_L = 10, 11.25, 12.5, 13.75, 15
% h_R = 0, 0.66, 1.32, 2
% theta = 0, 1, 2, 3, 4

\subsection{Full-Order Model}

%\subsubsection{Semi-Discrete Formulation}

We discretize equations \eqref{eq:2d_hb_conservative}--\eqref{eq:flux_vectors_2d} using a finite volume method on a Cartesian mesh with uniform spacings $\dx = L_x/(\Nx - 2)$ and $\dy = L_y/(\Ny - 2)$, where $\Nx$ and $\Ny$ denote the total number of cells in the $x$- and $y$-directions, respectively, including ghost cells used for the implementation of boundary conditions.

The discretization in space by integrating equation~\eqref{eq:2d_hb_conservative} over each cell and applying the divergence theorem:
\begin{equation}
\frac{d\bu_{i,j}}{dt} = -\frac{1}{\dx}\left(\hat{\blf}_{i+1/2,j} - \hat{\blf}_{i-1/2,j}\right) - \frac{1}{\dy}\left(\hat{\bg}_{i,j+1/2} - \hat{\bg}_{i,j-1/2}\right) + \bs_{g,i,j} + \bs_{f,i,j},
\label{eq:semidiscrete}
\end{equation}
where $\hat{\blf}_{i+1/2,j}$ and $\hat{\bg}_{i,j+1/2}$ denote numerical fluxes at cell interfaces.

%\subsubsection{Spatial Discretization: Rusanov Flux}

To approximate the fluxes we employ the Rusanov (local Lax--Friedrichs) flux~\cite{rusanov1962calculation}:
\begin{equation}
\hat{\blf}_{i+1/2,j} = \frac{1}{2}\left(\blf_{i,j} + \blf_{i+1,j}\right) - \frac{s_{\max,x}}{2}\left(\bu_{i+1,j} - \bu_{i,j}\right).
\label{eq:rusanov_flux_x}
\end{equation}
The local wave speed is defined by 
\begin{equation}
s_{\max,x} = \max\left(|u_{i,j}| + c_{i,j}, |u_{i+1,j}| + c_{i+1,j}\right),
\label{eq:wave_speed_x}
\end{equation}
with $c = \sqrt{gh\cos\thet}$ representing the fluid wave speed. An analogous formula applies for the $\bg$-fluxes with 
\begin{equation}
s_{\max,y} = \max(|v_{i,j}| + c_{i,j}, |v_{i,j+1}| + c_{i,j+1}).
\label{eq:wave_speed_y}
\end{equation}
Spatial derivatives of $h$ in the source terms are approximated using second-order centered differences.
% \begin{equation}
% \left.\frac{\partial h}{\partial x}\right|_{i,j} \approx \frac{h_{i+1,j} - h_{i-1,j}}{2\dx}, \quad 
% \left.\frac{\partial h}{\partial y}\right|_{i,j} \approx \frac{h_{i,j+1} - h_{i,j-1}}{2\dy}.
% \label{eq:centered_differences}
% \end{equation}

%\subsubsection{Operator Splitting Strategy}
For the time integration, we apply an operator splitting approach. 
Note that the friction source term $\bs_f$ renders the system stiff when the flow depth is small. 
This motivates the decomposition of \eqref{eq:semidiscrete} into a hyperbolic part (flux and gravity) and a friction part:
\begin{equation}
\frac{d\bu}{dt} = \mathcal{L}(\bu) + \bs_f(\bu),
\label{eq:operator_split}
\end{equation}
with
\begin{equation}
\mathcal{L}(\bu) = -\frac{1}{\dx}\left(\hat{\blf}_{i+1/2,j} - \hat{\blf}_{i-1/2,j}\right) - \frac{1}{\dy}\left(\hat{\bg}_{i,j+1/2} - \hat{\bg}_{i,j-1/2}\right) + \bs_g.
\label{eq:hyperbolic_operator}
\end{equation}
To advance the hyperbolic operator $\mathcal{L}(\bu)$, we use Heun's method (RK2), consisting of
\begin{align}
\text{Step 1 (Predictor):}&\quad \tilde{\bu}^{n+1} = \bu^n + \dt \, \mathcal{L}(\bu^n),
\label{eq:heun_step1}\\ 
\text{Step 2 (Corrector):}&\quad \bu^{n+1,*} = \bu^n + \frac{\dt}{2}\left[\mathcal{L}(\bu^n) + \mathcal{L}(\tilde{\bu}^{n+1})\right].
\label{eq:heun_step2}
\end{align}
After completing the explicit hyperbolic update~\eqref{eq:heun_step2}, the friction source term is treated semi-implicitly. For each discharge component,
\begin{equation}
q_k^{n+1} = \frac{q_k^{n+1,*}}{1 + \dt \, \alpha_k(h^{n+1,*}, q_k^{n+1,*})}, \quad k \in \{x,y\},
\label{eq:friction_update}
\end{equation}
where $\alpha_k$ is the friction coefficient defined in \eqref{eq:basal_shear_stress}. The flow depth remains unchanged, i.e., $h^{n+1} = h^{n+1,*}$.

\paragraph{Numerical Parameters.}
Given the final time $T = 30.0$\,s, the FOM simulations are conducted with the following numerical parameters:
\begin{itemize}
\item Spatial discretization: $\Nx \times \Ny = 102 \times 102$ (including ghost cells)\\[-5ex]
\item Grid spacing: $\dx = \dy = 4.0$\,m\\[-5ex]
\item Time step: $\dt = 0.005$\,s (fixed)\\[-5ex]
\item Snapshot collection interval: every 50 time steps ($\Delta t_{\text{snap}} = 0.25$\,s)\\[-5ex]
\item Number of temporal snapshots: $N_T = 122$
\end{itemize}

%%%%%%%%%%%%%%%%%%%%%%%%%%%%%%%%%%%
% Section 3
% TENSOR ROM FRAMEWORK
%%%%%%%%%%%%%%%%%%%%%%%%%%%%%%%%%%%
\section{Tensor-Based Reduced-Order Model}
\label{sec:trom}

The solution of the full-order model depends on several parameters: 
the yield stress, the initial upstream and downstream fluid depths, and the slope angle. 
We organize these parameters in a vector
\begin{equation}
\bmu = (\tauc, \hL, \hR, \thet) \in \mathcal{P} \subset \mathbb{R}^4.
\label{eq:parameter_vector}
\end{equation}
The purpose of the ROM is to provide a fast surrogate model capable of recovering the solution for any parameter value in the parameter domain $\mathcal{P}$ defined in \eqref{eq:paramDomain}. 
For this purpose, we approximate the parameter domain by a mesh, defined as the Cartesian product of nodes distributed over each parameter interval:
\begin{equation}
\label{paramsrange}
\begin{aligned}
\mathcal{P}_{\tauc} &= \{\tauc^{(1)}, \ldots, \tauc^{(\Ntau)}\} \subset [0, 6000]\,\text{Pa}, &
\mathcal{P}_{\hL} &= \{\hL^{(1)}, \ldots, \hL^{(\NL)}\} \subset [10, 15]\,\text{m}, \\
\mathcal{P}_{\hR} &= \{\hR^{(1)}, \ldots, \hR^{(\NR)}\} \subset [0, 2]\,\text{m}, &
\mathcal{P}_{\thet} &= \{\thet^{(1)}, \ldots, \thet^{(\Ntheta)}\} \subset [0^\circ, 4^\circ].
\end{aligned}
\end{equation}
The parameter mesh is then given by
\[
\widehat{\mathcal{P}} = \mathcal{P}_{\tauc} \times \mathcal{P}_{\hL} \times \mathcal{P}_{\hR} \times \mathcal{P}_{\thet}.
\]

The offline stage of the ROM proceeds by assembling a sixth-order snapshot tensor for each field variable $\phi \in \{h, q_x, q_y\}$:
\begin{equation}
\QQ_\phi \in \mathbb{R}^{N_x N_y \times \Ntau \times \NL \times \NR \times \Ntheta \times N_T},
\label{eq:snapshot_tensor}
\end{equation}
where the first mode corresponds to the spatial discretization ($N_x N_y = 100 \times 100 = 10^4$), modes 2--5 correspond to the parameter dimensions ($\tauc, \hL, \hR, \thet$), and the last mode corresponds to time ($N_T = 122$). 
The tensor entries are populated by running FOM simulations:
\begin{equation}
[\QQ_\phi]_{i, j_1, j_2, j_3, j_4, k} = \phi(\mathbf{x}_i, t_k; \tauc^{(j_1)}, \hL^{(j_2)}, \hR^{(j_3)}, \thet^{(j_4)}).
\label{eq:tensor_population}
\end{equation}

\subsection{Offline order reduction}

Following~\cite{mamonov2022interpolatory,mamonov2024tensorial}, we apply a low-rank tensor decomposition for model reduction. In this paper, we employ the (truncated) higher-order SVD~\cite{de2000multilinear}, which provides a quasi-optimal approximation of the snapshot tensor in Tucker format:
\begin{equation}
\QQ_\phi \approx \mathcal{G}_\phi \times_1 \bW_\phi \times_2 \Sig_\phi^{(1)} \times_3 \Sig_\phi^{(2)} \times_4 \Sig_\phi^{(3)} \times_5 \Sig_\phi^{(4)} \times_6 \bV_\phi,
\label{eq:tucker_form}
\end{equation}
where
\begin{itemize}
\item $\mathcal{G}_\phi \in \mathbb{R}^{\tilde{N}_x \times \tilde{N}_{\tauc} \times \tilde{N}_L \times \tilde{N}_R \times \tilde{N}_\theta \times \tilde{N}_T}$ is the core tensor,\\[-5ex]
\item $\bW_\phi \in \mathbb{R}^{N_x N_y \times \tilde{N}_x}$ is the spatial factor matrix,\\[-5ex]
\item $\Sig_\phi^{(k)} \in \mathbb{R}^{N_k \times \tilde{N}_k}$ are the parameter factor matrices, $k = 1,2,3,4$,\\[-5ex]
\item $\bV_\phi \in \mathbb{R}^{N_T \times \tilde{N}_T}$ is the temporal factor matrix.
\end{itemize}

The multilinear ranks $\tilde{N}_x, \tilde{N}_{\tauc}, \tilde{N}_L, \tilde{N}_R, \tilde{N}_\theta, \tilde{N}_T$ are selected using an energy criterion:
\begin{equation}
\tilde{N}_k = \min \left\{ r : \frac{\sum_{i=1}^r \sigma_{i,k}}{\sum_{i=1}^{N_k} \sigma_{i,k}} \geq 1 - \epsilon \right\},
\label{eq:rank_selection}
\end{equation}
where $\sigma_{i,k}$ are the singular values of the $k$-mode unfolding of $\QQ_\phi$. 
We use $\epsilon = 10^{-3}$ as a threshold for all modes.
The multilinear ranks obtained from the truncated HOSVD using the energy criterion above are
\[
(\tilde{N}_x, \tilde{N}_{\tauc}, \tilde{N}_L, \tilde{N}_R, \tilde{N}_\theta, \tilde{N}_T)
= (5589,\, 11,\, 5,\, 4,\, 5,\, 66).
\]

\subsection{Online Phase: non-Intrusive Tensor ROM}\label{sec:3.2}

For the non-intrusive tensor ROM, we find an approximate solution for each new parameter value $\bmu^* = (\mu_1^*, \dots, \mu_4^*) = (\tauc^*, \hL^*, \hR^*, \thet^*)$
directly from the low-rank representation of the snapshot tensor. This is accomplished through several steps involving fast interpolation and low-dimensional linear algebra.

First, we compute interpolation coefficients for a cubic spline interpolation of an (abstract smooth) function at the point $\mu_k^*$
on the grid $\mathcal{P}_k$:
\begin{equation}\label{interp}
	\chi_k(\mu_k^*) = \text{interp. coeff.}(\mathcal{P}_k,  \mu_k^*), \quad k = 1,2,3,4.
\end{equation}
The vector $\chi_k \in \mathbb{R}^{N_k}$ contains the interpolation coefficients corresponding to the parameter value $\mu_k^*$ on the grid $\mathcal{P}_k$. 
Next, the Tucker tensor from \eqref{eq:tucker_form}  is contracted along all parameter modes with the interpolation vectors:
\begin{equation}
	\Ccore_\phi(\bmu^*) = \mathcal{G}_\phi \times_2 \left(\Sig_\phi^{(1)}\chi_1(\mu_1^*)\right) \times_3 \dots \times_5 \left(\Sig_\phi^{(4)}\chi_4(\mu_4^*)\right).
\end{equation}
This yields the low-dimensional matrix $\Ccore_\phi(\bmu^*) \in \mathbb{R}^{\tilde{N}_x \times \tilde{N}_T}$.
The  solution  is then reconstructed via
\begin{equation}
	\bX_\phi(\bmu^*) =\bW_\phi \Ccore_\phi(\bmu^*) {\bV}_\phi^T  \in \mathbb{R}^{N_x N_y \times N_T}.
	\label{eq:matrix_reconstruction}
\end{equation}
The $k$th column of $\bX_\phi(\bmu^*)$ approximates $\phi(\mathbf{x}, t_k; \bmu^*)$.

\begin{remark}\rm
	The rationale behind the solution recovery in \eqref{interp}--\eqref{eq:matrix_reconstruction} can be understood by considering the case where $\bmu^*$ coincides with a grid point in $\widehat{\mathcal{P}}$, i.e., a parameter comes from the training set. In this case, the interpolation coefficient vectors $\chi_k(\mu_k^*)$ contain all zeros except for a single entry equal to $1$. Let $\epsilon = 0$ in \eqref{eq:rank_selection}, so that the LRTD in \eqref{eq:tucker_form} is exact. Then straightforward calculations show that the tensor ROM recovers the FOM solution for $\bmu^*$ \emph{exactly}. 
	
	For $\bmu^*$ \emph{not} in the training set, the TROM performs local interpolation to reconstruct the solution up to interpolation errors, while the offline LRTD enables efficient online computations using low-dimensional quantities.
\end{remark}

An additional dimension reduction can be done by computing the SVD of the small-size core matrix:
\begin{equation}
	\Ccore_\phi(\bmu^*) = \widehat{\bU}_\phi \widehat{\Sig}_\phi \widehat{\bV}_\phi^T.
\end{equation}
and truncating it to the first $\ell_\phi$ singular values $\hat{\sigma}_i$ of $\Ccore_\phi(\bmu^*)$, where the local reduced dimension is determined from:
\begin{equation}
	\ell_\phi = \min\left\{k : \frac{\sum_{i=1}^k \hat{\sigma}_i}{\sum_{i=1}^{\min(\tilde{N}_x,\tilde{N}_T)} \hat{\sigma}_i} \geq 1 - \epsloc \right\}.
\end{equation}
In our experiments, we set $\epsloc = 10^{-3}$ and replace $\Ccore_\phi(\bmu^*) $ in \eqref{eq:matrix_reconstruction} with its truncated SVD.

\begin{remark}[Encoder and Decoder]\rm  \label{rem2}
	The non-intrusive online procedure here differs from the interpolatory TROM in \cite{mamonov2022interpolatory}, where a parameter-specific (local) spatial basis is recovered and the governing equations, i.e. eq.~\eqref{eq:2d_hb_conservative} in our case, and the initial conditions are then projected onto the local TROM space and integrated forward in time. 
	
	The non-intrusive TROM does not invoke any time integration or governing equation, and it is more in the spirit of an encoder--decoder scheme, where the encoder is given by the (nonlinear) mapping:
	\[
	\bmu^* \to \{\Sig_\phi^{(k)}\chi_k(\mu_k^*)\}_{k=1,\dots,4},
	\]  
	where the reduced dimensions of the vectors $\Sig_\phi^{(k)}$ and $\chi_1(\mu_k^*)$ are given by the multilinear tensor ranks $\tilde N_k$. 
	
	The decoder is given by the (multilinear) mapping 
	\[
	\{\Sig_\phi^{(k)}\chi_k(\mu_k^*)\}_{k=1,\dots,4} \to \bX_\phi(\bmu^*).
	\] 
	Both the encoder and decoder are ``learned'' by performing HOSVD on the snapshot tensors. Unlike some NN-based encoder--decoder approaches, the one discussed in this paper should be amenable to rigorous error analysis for solutions smoothly depending on parameters; see~\cite{mamonov2025priori} for arguments in the context of intrusive (projection-based) TROMs. We also do not see how a similar non-intrusive encoder--decoder scheme can be realized for parametric solutions within more traditional POD or Reduced Basis frameworks.   
\end{remark}

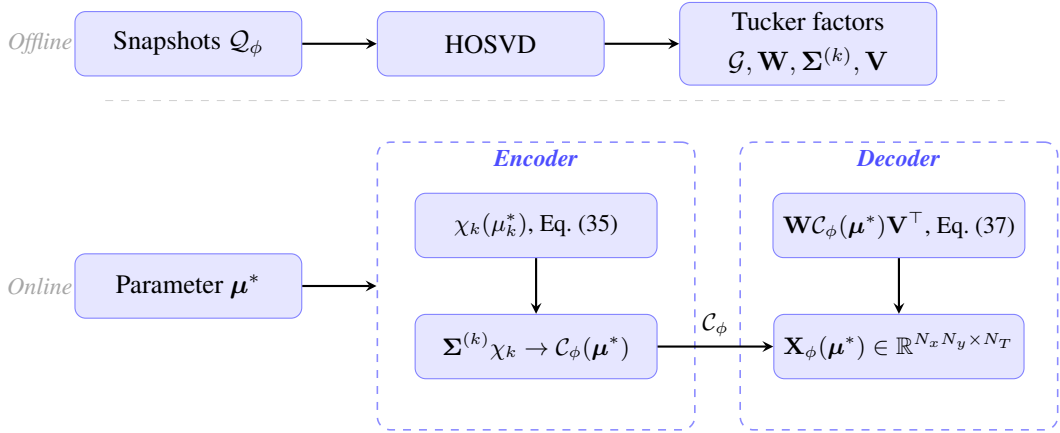
\begin{figure}[H]
\centering
\begin{tikzpicture}[
  box/.style={rectangle, rounded corners=4pt, minimum width=3.0cm,
              minimum height=0.85cm, align=center, font=\small},
  arr/.style={->, >=stealth, thick},
  every node/.style={font=\small}
]

% ---- OFFLINE ROW (y=0) ----
\node[box, fill=blue!10, draw=blue!50]
  (snap) at (0,0) {Snapshots $\mathcal{Q}_\phi$};

\node[box, fill=blue!10, draw=blue!50]
  (hosvd) at (4.0,0) {HOSVD};

\node[box, fill=blue!10, draw=blue!50, minimum width=3.4cm]
  (factors) at (8.2,0) {Tucker factors\\[3pt]
    $\mathcal{G},\mathbf{W},\boldsymbol{\Sigma}^{(k)},\mathbf{V}$};

\draw[arr] (snap.east)  -- (hosvd.west);
\draw[arr] (hosvd.east) -- (factors.west);

\node[font=\footnotesize\itshape, gray!70, anchor=east]
  at (-1.4, 0)   {Offline};
\node[font=\footnotesize\itshape, gray!70, anchor=east]
  at (-1.4,-3.2) {Online};

% ---- DIVIDER ----
\draw[gray!35, dashed] (-1.1,-0.75) -- (10.8,-0.75);

% ---- INPUT ----
\node[box, fill=blue!10, draw=blue!50]
  (mu) at (0,-3.2) {Parameter $\boldsymbol{\mu}^*$};

% ---- ENCODER inner dashed box ----
\node[draw=blue!45, rounded corners=5pt, dashed, line width=0.6pt,
      minimum width=4.2cm, minimum height=3.8cm]
  (encbox) at (4.6,-3.2) {};

\node[font=\footnotesize\bfseries, blue!70]
  at (4.6,-1.5) {\textit{Encoder}};

\node[box, fill=blue!10, draw=blue!50, minimum width=3.2cm]
  (interp) at (4.6,-2.4)
  {\footnotesize $\chi_k(\mu_k^*)$, Eq.~\eqref{interp}};

\node[box, fill=blue!10, draw=blue!50, minimum width=3.2cm]
  (contract) at (4.6,-4.0)
  {\footnotesize $\boldsymbol{\Sigma}^{(k)}\chi_k
    \to \mathcal{C}_\phi(\boldsymbol{\mu}^*)$};

\draw[arr] (interp.south) -- (contract.north);

% ---- DECODER inner dashed box ----
\node[draw=blue!45, rounded corners=5pt, dashed, line width=0.6pt,
      minimum width=4.2cm, minimum height=3.8cm]
  (decbox) at (9.4,-3.2) {};

\node[font=\footnotesize\bfseries, blue!70]
  at (9.4,-1.5) {\textit{Decoder}};

\node[box, fill=blue!10, draw=blue!50, minimum width=3.2cm]
  (decode) at (9.4,-2.4)
  {\footnotesize $\mathbf{W}\mathcal{C}_\phi(\boldsymbol{\mu}^*)
    \mathbf{V}^\top$, Eq.~\eqref{eq:matrix_reconstruction}};

\node[box, fill=blue!10, draw=blue!50, minimum width=3.2cm]
  (sol) at (9.4,-4.0)
  {\footnotesize $\mathbf{X}_\phi(\boldsymbol{\mu}^*)
    \in\mathbb{R}^{N_xN_y\times N_T}$};

\draw[arr] (decode.south) -- (sol.north);

% ---- CONNECTIONS ----
\draw[arr] (mu.east) -- (encbox.west |- mu.east);

\draw[arr] (contract.east) -- (sol.west |- contract.east)
  node[midway, above, font=\footnotesize] {$\mathcal{C}_\phi$};

\end{tikzpicture}
\caption{Encoder--decoder interpretation of the NI-TROM.
The offline stage compresses the snapshot tensor via HOSVD.
The online encoder maps $\boldsymbol{\mu}^*$ to the latent
matrix $\mathcal{C}_\phi(\boldsymbol{\mu}^*)$; the decoder
reconstructs $\mathbf{X}_\phi(\boldsymbol{\mu}^*)$ via
multilinear operations, without any PDE solve.}
\label{fig:enc_dec}
\end{figure}

%%%%%%%%%%%%%%%%%%%%%%%%%%%%%%%%%%%
% Section 4
%%%%%%%%%%%%%%%%%%%%%%%%%%%%%%%%%%%
\section{Numerical Results}
\label{sec:num}
In this section, we present numerical results for the non-intrusive TROM and assess its accuracy by comparison with the FOM. As follows from the description above, the non-intrusive TROM does not require time-stepping of a reduced model; instead, the solution is reconstructed directly from the low-rank tensor representation.

We present a systematic study of the TROM accuracy across a range of parameters.
Two complementary diagnostics are used throughout the paper --- 
(1) the flow front evolution in the $(x,y)$-plane and 
(2) the cross-sectional plug/shear structure along the channel
centerline at final time for the visco--plastic cases with $\tau_c > 0$.We report dimensional parameter values consistent with \eqref{paramsrange}, but omit units for brevity.

We examine the influence of parameters in the following regimes --- 
Section \ref{subsubsec:newtonian} presents the Newtonian case for different initial conditions (varying $h_L$ and $h_R$), 
Section \ref{subsubsec:tauc_impact} discusses the non-Newtonian case and the effect of increasing $\tauc$,
Section \ref{subsubsec:hR_impact} presents results for varying the wet-bed depth $h_R$ for a fixed value of the yield stress $\tau_c = 1100$ with particular emphasis on computing the yield surface, 
Section \ref{subsubsec:theta_impact} presents the effect of changing the bed slope $\theta$
for fixed values of $(h_L, h_R) = (11,1.2)$ and $\tau_c = 6000$.

In Appendix~\ref{subsec:long_fom_solution}, we present long-time FOM simulations in the regime
$\tauc=6000$, $\thet=0$, $\hL=11$, comparing the dry-bed ($\hR=0$) and wet-bed ($\hR=1.2$) cases.
These numerical results demonstrate that the presence of a wet bed significantly enhances the downstream propagation of the leading front, in agreement with recent results reported in \cite{muchiri2024numerical}.
In the wet-bed simulations, after an initial faster phase, the flow transitions into a slow creeping regime and eventually reaches the outflow boundary of the computational domain. Therefore, we do not report stopping times; instead, we show that the TROM captures the transition from the initial fast-moving phase to a slowly moving regime and accurately reproduces the propagation speed of the leading front.

%---------------------
\subsection{Newtonian Fluid ($\tauc = 0$)}
\label{subsubsec:newtonian}
%---------------------
Setting $\tauc = 0$ and $\thet = 0$ reduces the
Herschel--Bulkley model to a standard shallow-water system.
To illustrate the behavior of the Newtonian fluid with a flat bottom for different initial conditions, 
we consider two upstream depths $\hL =11, 13$ paired
with two bed configurations: a \emph{dry bed} $\hR = 0$ and
a \emph{wet bed} $\hR = 1.2$. Both values of $\hL$ considered here and $\hR=1.2$ are out-of-sample, i.e., these values are not included in the training dataset.

\begin{figure}[H]
  \centering
  \begin{subfigure}[t]{0.49\linewidth}
    \centering
    \includegraphics[width=\linewidth]{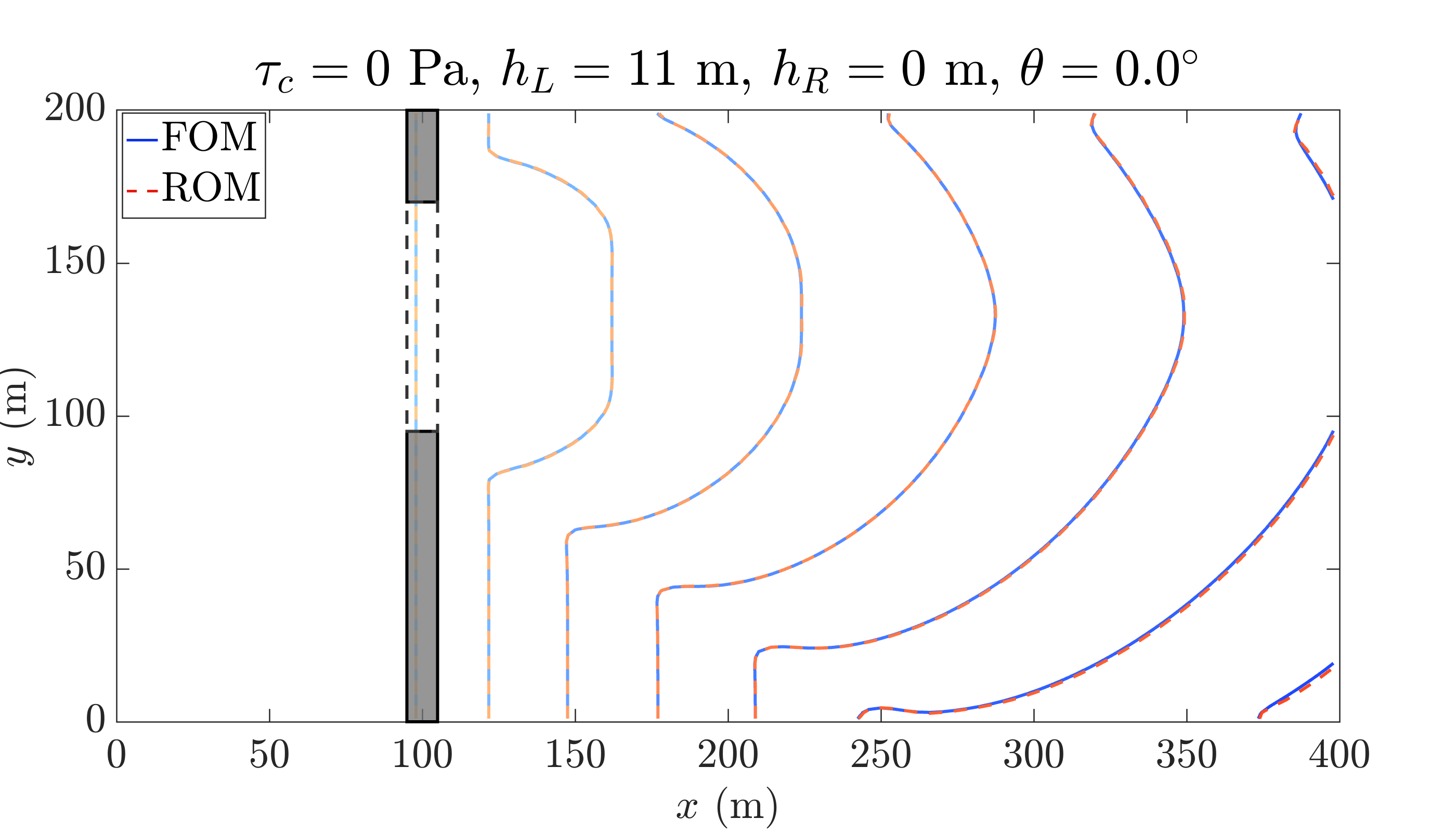}
    \caption{Dry bed ($\hR = 0$),
             $t = 3, 6, \ldots, 18$
             (front exits domain before $t = 21$).}
    \label{fig:front_dry_11}
  \end{subfigure}
  \hfill
  \begin{subfigure}[t]{0.49\linewidth}
    \centering
    \includegraphics[width=\linewidth]{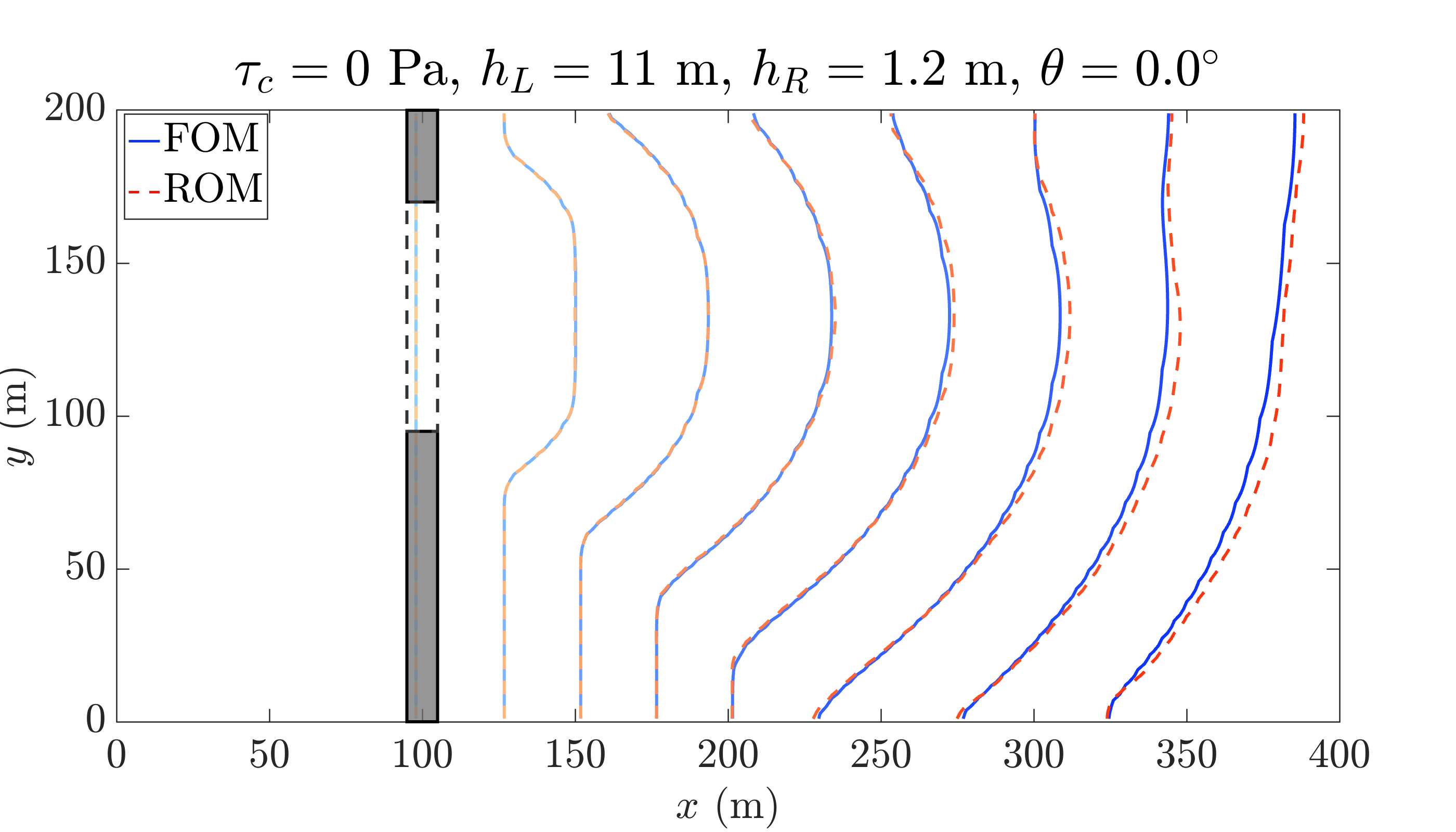}
    \caption{Wet bed ($\hR = 1.2$),
             $t = 3, 6, \ldots, 21$.}
    \label{fig:front_wet_11}
  \end{subfigure}
  \caption{Flow front evolution for the Newtonian fluid
           $\tauc = 0$ and parameters $\hL = 11$, $\thet = 0$.
           Blue solid: FOM; red dashed: NI-TROM.}
  \label{fig:front_newtonian_11}
\end{figure}

\begin{figure}[H]
  \centering
  \begin{subfigure}[t]{0.49\linewidth}
    \centering
    \includegraphics[width=\linewidth]{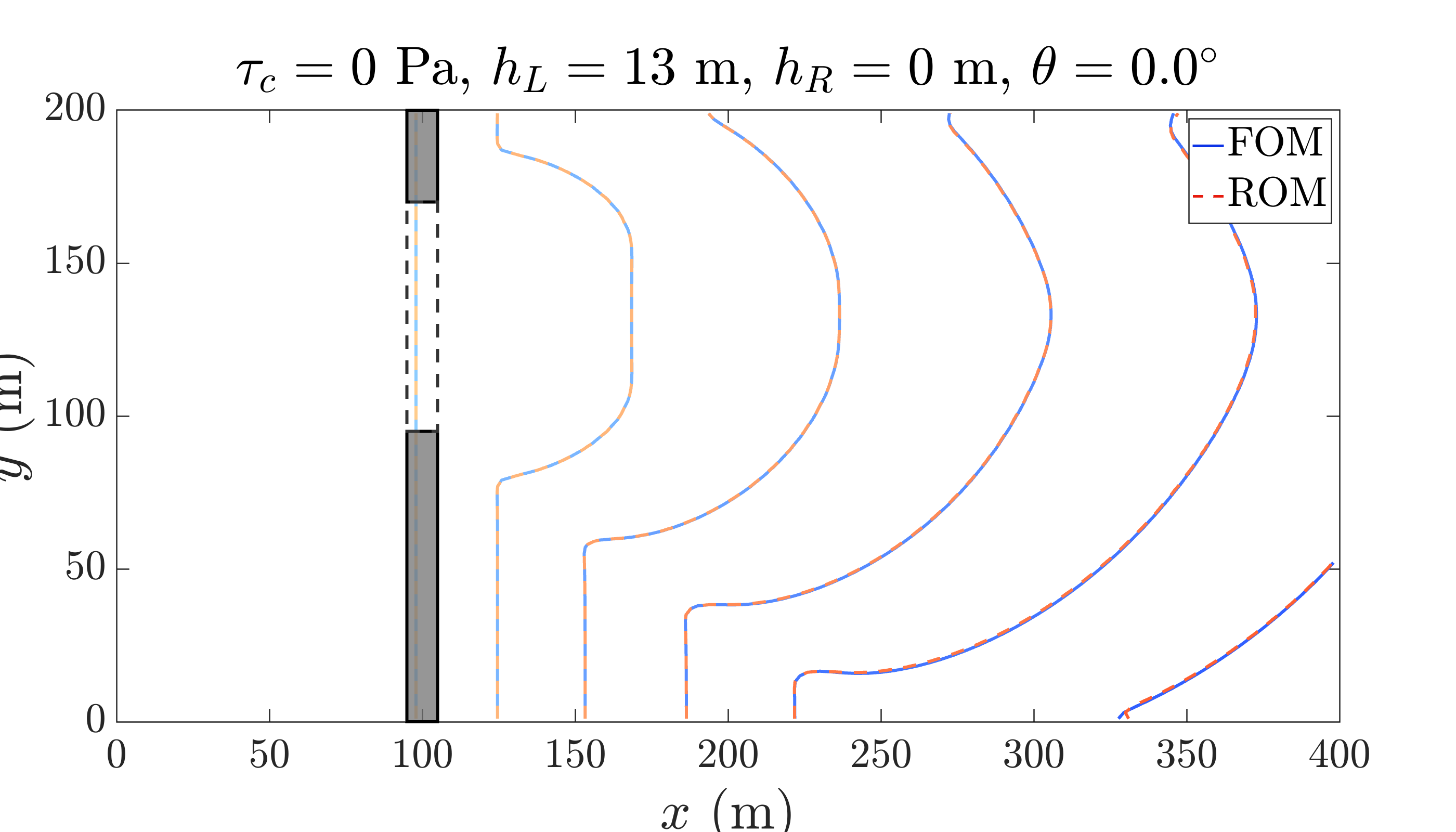}
    \caption{Dry bed ($\hR = 0$),
             $t = 3, 6, \ldots, 15$
             (front exits domain before $t = 18$ due to
             higher upstream head).}
    \label{fig:front_dry_13}
  \end{subfigure}
  \hfill
  \begin{subfigure}[t]{0.49\linewidth}
    \centering
    \includegraphics[width=\linewidth]{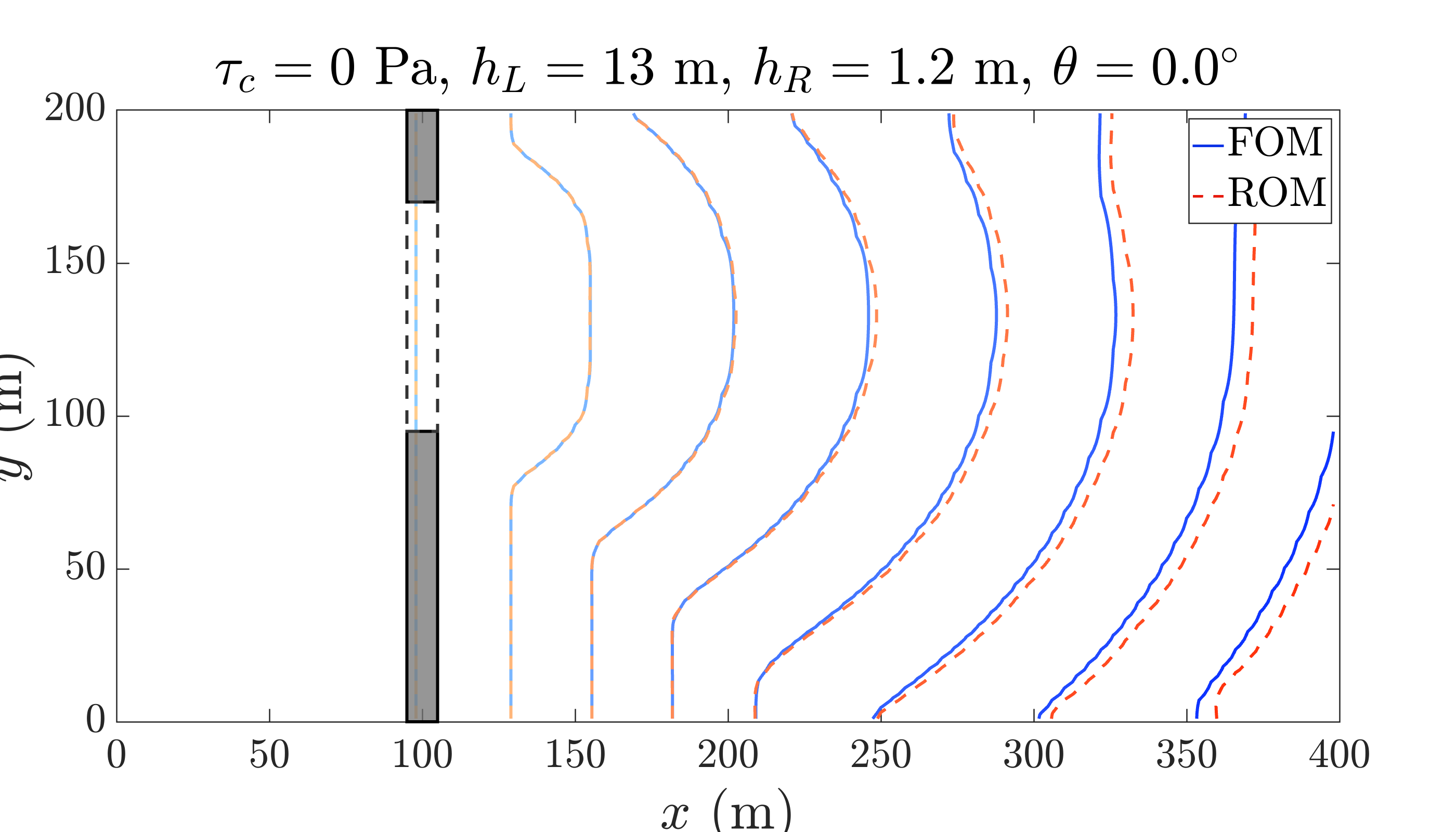}
    \caption{Wet bed ($\hR = 1.2$),
             $t = 3, 6, \ldots, 21$.}
    \label{fig:front_wet_13}
  \end{subfigure}
  \caption{Flow front evolution for the Newtonian fluid
           $\tauc = 0$ and parameters $\hL = 13$, $\thet = 0$.
           Blue solid: FOM; red dashed: NI-TROM.}
  \label{fig:front_newtonian_13}
\end{figure}

Next, front contours are extracted at equal intervals $\Delta t = 3$ and depicted in Figures~\ref{fig:front_newtonian_11} and~\ref{fig:front_newtonian_13}
for $\hL=11$ and $\hL=13$, respectively.
These Figures show that, on the dry bed, the flood wave encounters no ambient fluid and propagates unimpeded.
For $\hR = 0$, the front reaches the downstream boundary ($x = 400$) before $t = 21$; thus contours are shown only until $t=18$ and $t=15$ in Figs.~\ref{fig:front_dry_11} and~\ref{fig:front_dry_13}, respectively. Comparison of 
Figs.~\ref{fig:front_dry_11} and~\ref{fig:front_dry_13} also demonstrates that higher $\hL=13$ (Fig.~\ref{fig:front_dry_13}) leads to a faster acceleration of the fluid front. 
This is also true in wet-bed simulations with $\hR=1.2$, i.e., the leading wave propagates further for $\hL=13$ then $\hL=11$ (c.f. Figs.~\ref{fig:front_wet_13} and~\ref{fig:front_wet_11}).
In addition, for the wet bed, the ambient fluid layer exerts a backflow effect
that decelerates the wave, and all seven contours
($t = 3, 6, \ldots, 21$) remain within the domain for both
values of $\hL$
(Figs.~\ref{fig:front_wet_11} and~\ref{fig:front_wet_13}).
Therefore, comparing the dry- and wet-bed simulations, the front in wet-bed simulations is noticeably slower,
reflecting the additional resistance from the downstream fluid.

Simulations in this section demonstrate that our non-intrusive TROM model performs very well in the Newtonian regime. In particular, our reduced model accurately reproduces fully-resolved dam-break simulations, including the propagation of the leading front. 

% ------------------------
\subsection{Non-Newtonian Fluid ($\tauc > 0$)}
\label{subsubsec:tauc_impact}
% ------------------------
In this section, we investigate the performance of the TROM in the non-Newtonian regime. We start by varying the yield stress $\tauc$.
In particular, we fix out-of-sample values of parameters 
$\hL = 11$, $\hR = 1.2$, $\thet = 3.2$ and consider 
$\tauc \in \{300,\,1100,\,3000,\, 6000\}$ to analyze the
effect of viscoplasticity.
Here, $\tauc=300, 1100$ are also  out-of-sample values, while $\tauc=3000, 6000$ are in-sample values. Thus, all four regimes can be considered out-of-sample regimes in the full parameter domain
$\widehat{\mathcal{P}}$.

\begin{figure}[H]
  \centering
  \begin{subfigure}[b]{0.49\linewidth}
    \centering
    \includegraphics[width=\linewidth]{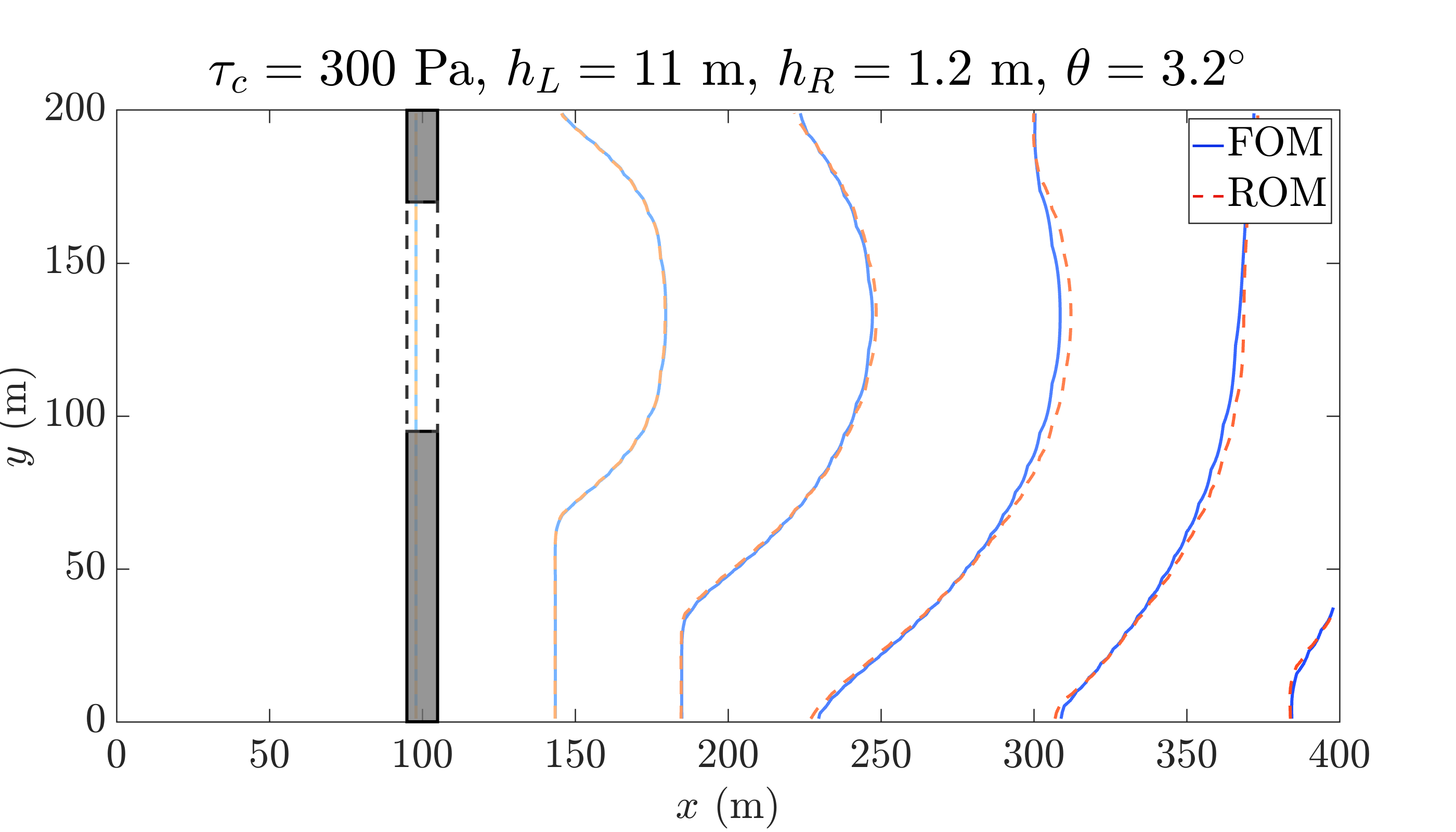}
    \caption{Flow front evolution, $t = 0, 5, \ldots, 25$.}
    \label{fig:front_tau300}
  \end{subfigure}
  \hfill
  \begin{subfigure}[b]{0.49\linewidth}
    \centering
    \includegraphics[width=\linewidth]{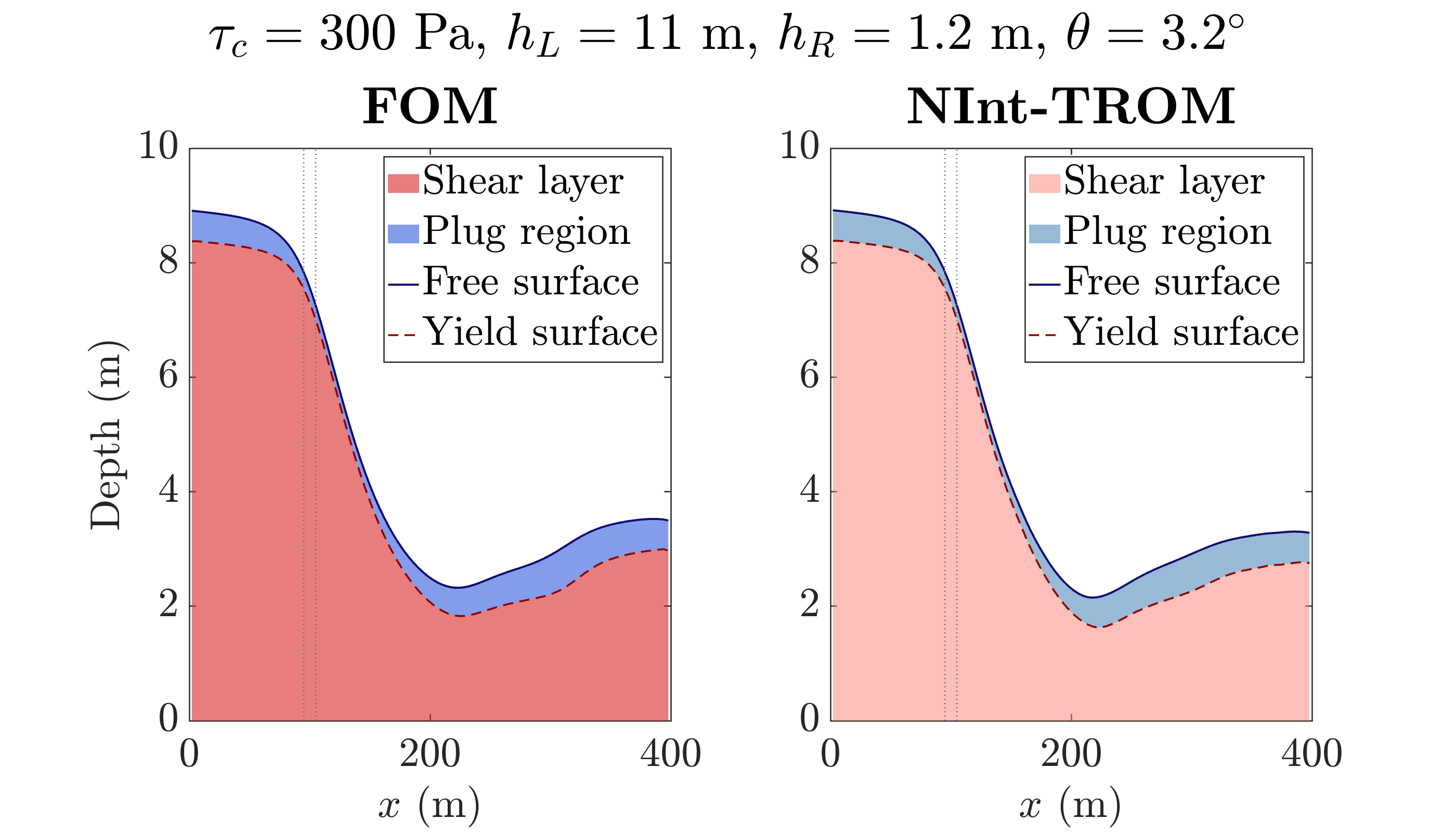}
    \caption{Flow front evolution, $t = 0, 5, \ldots, 30$.}
    \label{fig:xs_tau300}
  \end{subfigure}
  \caption{$\tauc = 300$, $\hL = 11$, $\hR = 1.2$,
           $\thet = 3.2^{\circ}$.
           Blue solid: FOM; red dashed: NI-TROM.}
  \label{fig:tau300}
\end{figure}

\begin{figure}[H]
  \centering
  \begin{subfigure}[b]{0.49\linewidth}
    \centering
    \includegraphics[width=\linewidth]{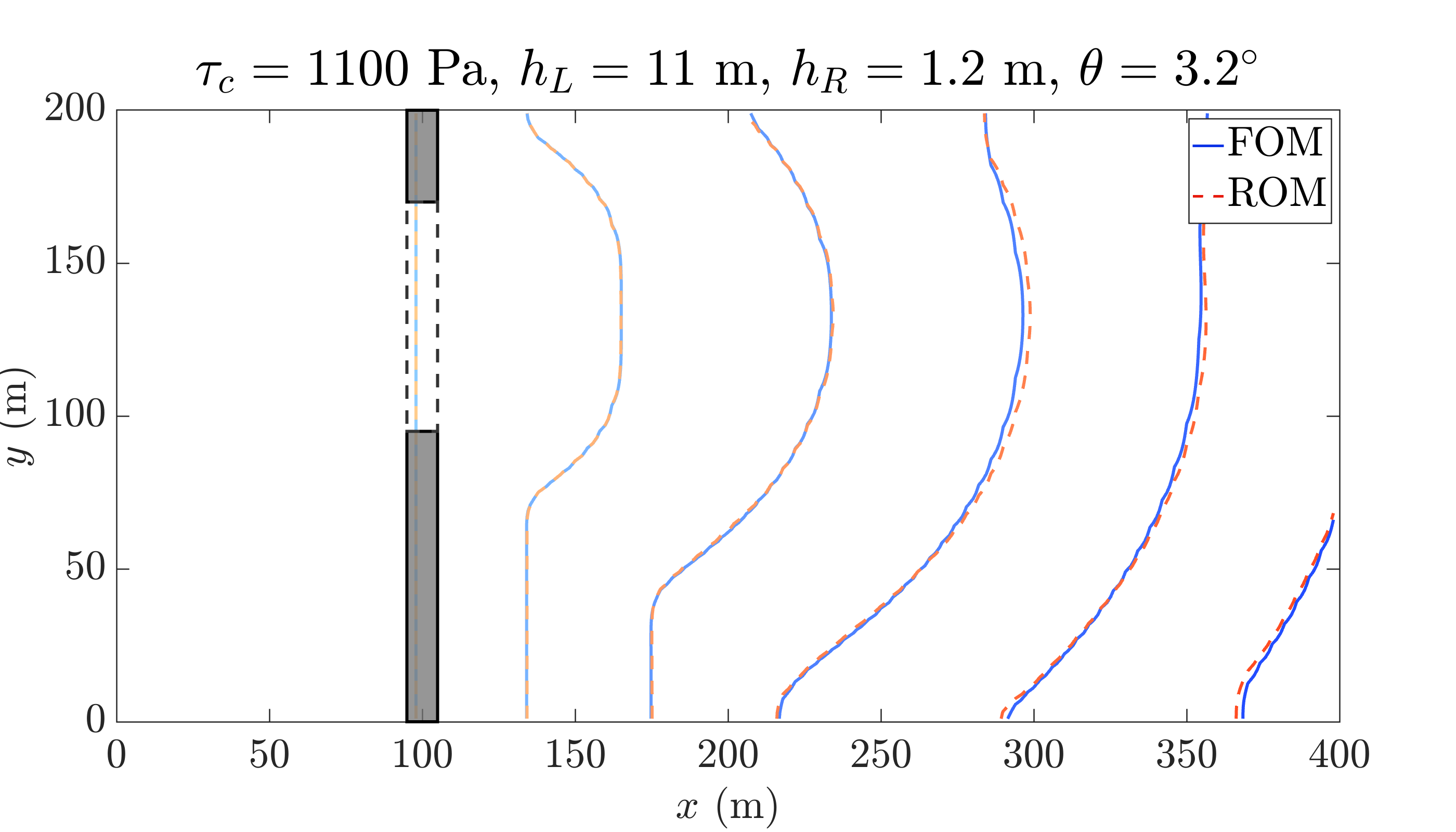}
    \caption{Flow front evolution, $t = 0, 5, \ldots, 25$.}
    \label{fig:front_tau1100}
  \end{subfigure}
  \hfill
  \begin{subfigure}[b]{0.49\linewidth}
    \centering
    \includegraphics[width=\linewidth]{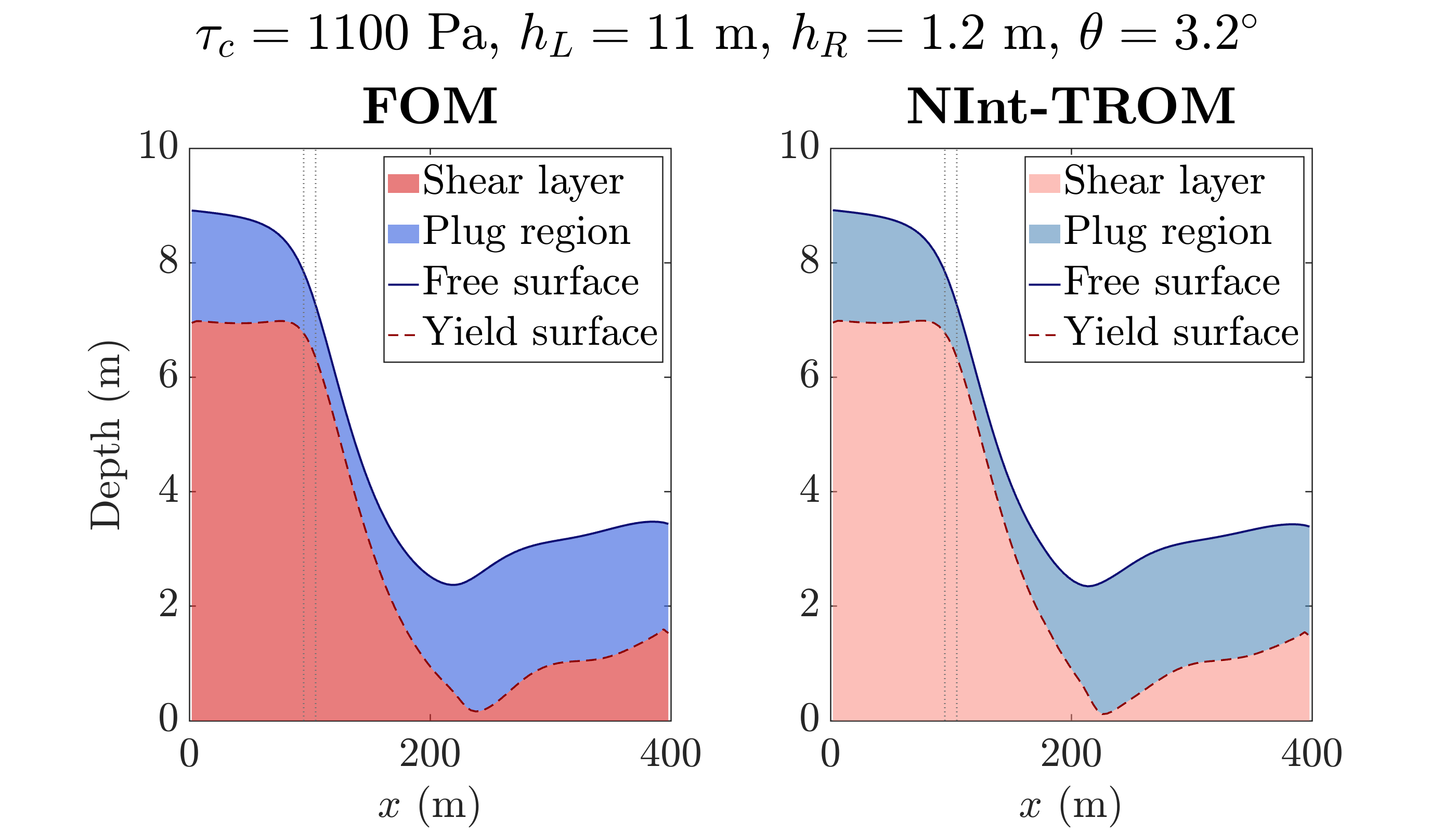}
    \caption{Flow front evolution, $t = 0, 5, \ldots, 30$.}
    \label{fig:xs_tau1100}
  \end{subfigure}
  \caption{$\tauc = 1100$, $\hL = 11$, $\hR = 1.2$,
           $\thet = 3.2^{\circ}$.
           Blue solid: FOM; red dashed: NI-TROM.}
  \label{fig:tau1100}
\end{figure}

\begin{figure}[H]
  \centering
  \begin{subfigure}[b]{0.49\linewidth}
    \centering
    \includegraphics[width=\linewidth]{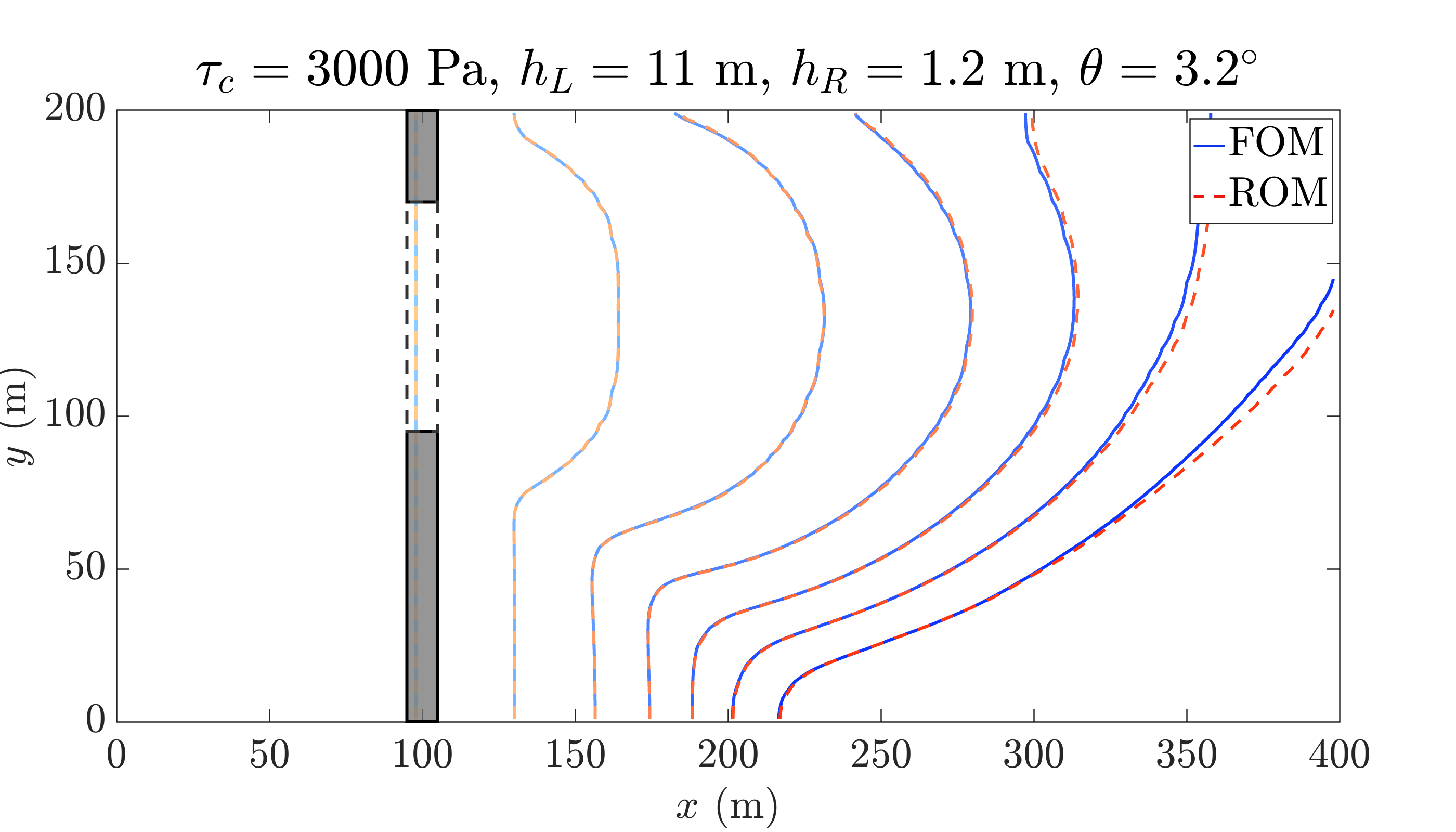}
    \caption{Flow front evolution, $t = 0, 5, \ldots, 30$.}
    \label{fig:front_tau3000}
  \end{subfigure}
  \hfill
  \begin{subfigure}[b]{0.49\linewidth}
    \centering
    \includegraphics[width=\linewidth]{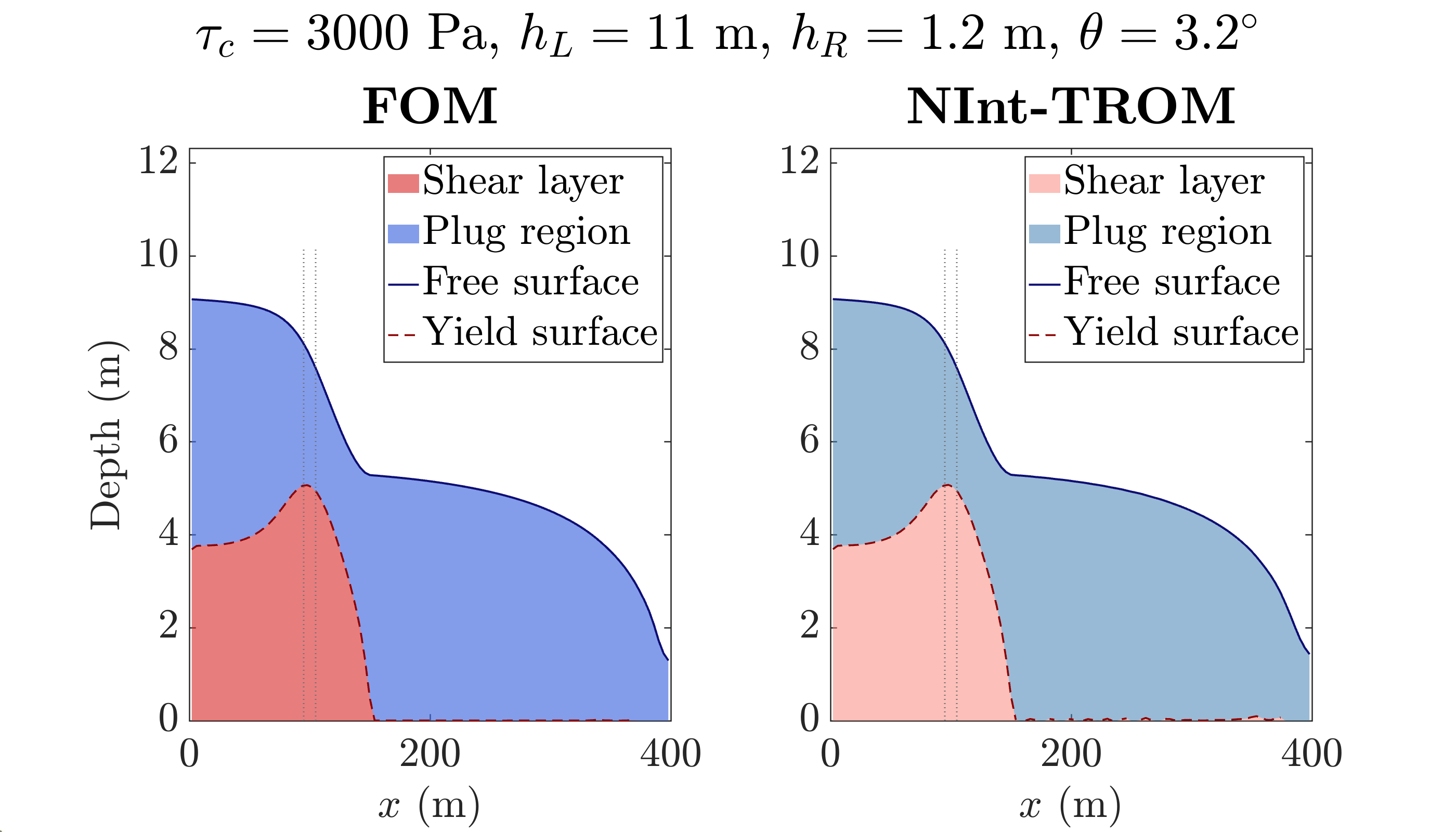}
    \caption{Cross-sectional plug/shear structure at $T = 30$.}
    \label{fig:xs_tau3000}
  \end{subfigure}
  \caption{$\tauc = 3000$, $\hL = 11$, $\hR = 1.2$,
           $\thet = 3.2^{\circ}$.
           Blue solid: FOM; red dashed: NI-TROM.}
  \label{fig:tau3000}
\end{figure}

\begin{figure}[H]
  \centering
  \begin{subfigure}[b]{0.49\linewidth}
    \centering
    \includegraphics[width=\linewidth]{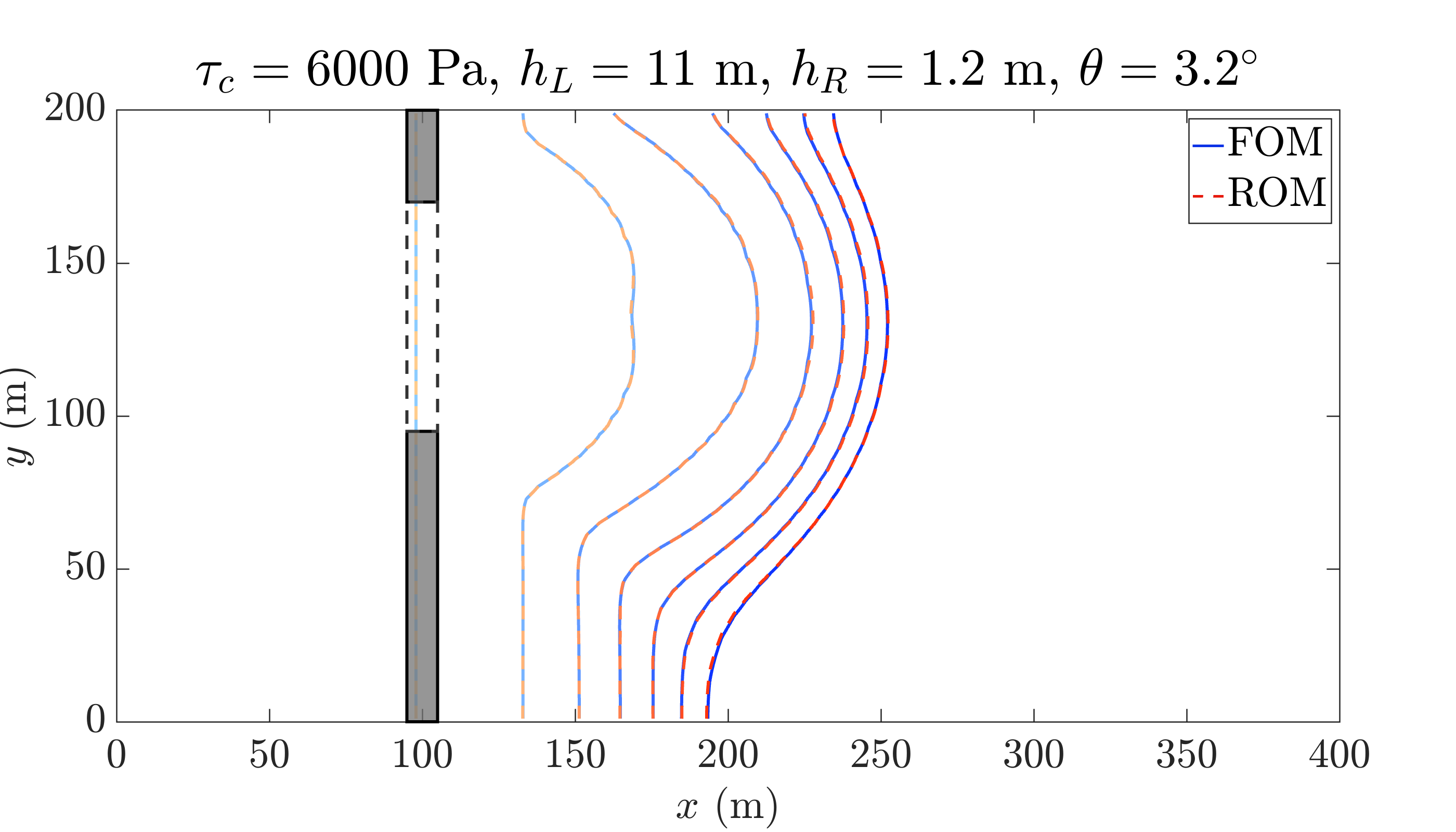}
    \caption{Flow front evolution, $t = 0, 5, \ldots, 30$.}
    \label{fig:front_tau6000}
  \end{subfigure}
  \hfill
  \begin{subfigure}[b]{0.49\linewidth}
    \centering
    \includegraphics[width=\linewidth]{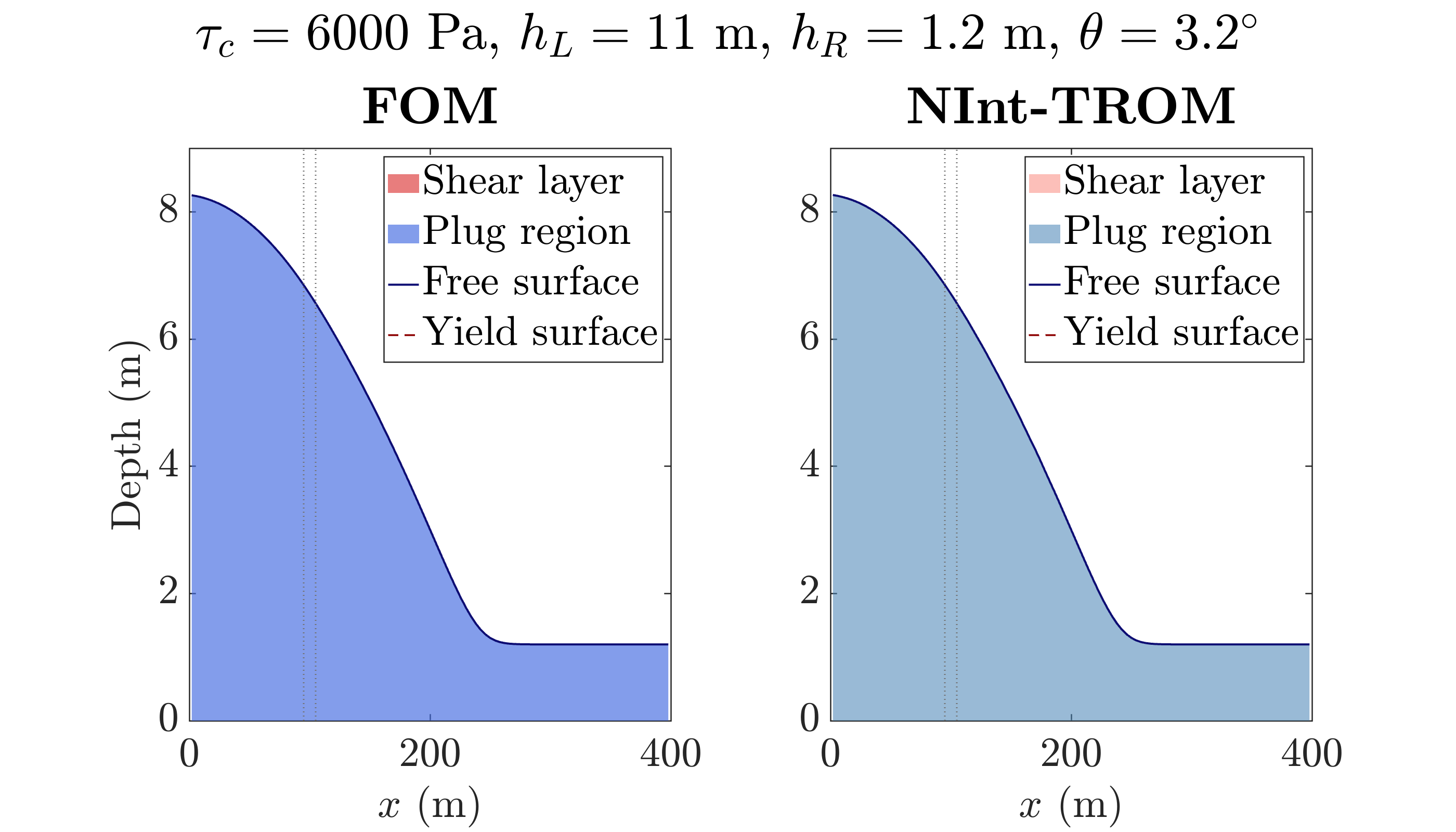}
    \caption{Cross-sectional plug/shear structure at $T = 30$.}
    \label{fig:xs_tau6000}
  \end{subfigure}
  \caption{$\tauc = 6000$, $\hL = 11$, $\hR = 1.2$,
           $\thet = 3.2^{\circ}$.
           Blue solid: FOM; red dashed: NI-TROM.}
  \label{fig:tau6000}
\end{figure}

Similar to the previous section, leading front contours with the interval $\Delta t=6$
are plotted 
Figures \ref{fig:front_tau300}--
\ref{fig:front_tau6000}.
In addition, Figures 
\ref{fig:xs_tau300}--
\ref{fig:xs_tau6000}
also depict 
cross-sections of the
plug/shear regions along the centerline $y=100$ at the final time $T = 30$.

%As $\tauc$ increases from 300 to 6000, the yield stress provides progressively stronger resistance to deformation, and this is reflected directly in the front positions (Figs.~\ref{fig:front_tau300}--\ref{fig:front_tau6000}).

At $\tauc = 300$ and $\tauc=1100$ (Fig.~\ref{fig:front_tau300} and~\ref{fig:front_tau1100}) the non-Newtonian effects remain weak. The wave propagates rapidly with an asymmetric front shape, characteristic of a concentrated jet issuing from the breach. 
The leading front reaches the
downstream boundary before $t = 30$ in both cases, and only the contours and only contours until $t=25$ are shown. We observe a weak non-Newtonian effect where the leading front slows down slightly for a larger value of $\tauc=1100$. Our TROM correctly predicts the propagation of the leading front as well as the plug/shear regions. 
At $\tauc = 3000$ and $\tauc = 6000$ 
(Figs.~\ref{fig:front_tau3000} and~\ref{fig:front_tau6000}) visco--plastic effects become more pronounced.
%
% The contours become more elongated near the breach and progressively rounded farther downstream, indicating that the yield stress suppresses lateral spreading relative to the low-$\tauc$ case.
%
For $\tauc = 6000$ (Fig.~\ref{fig:front_tau6000}), the fluid is strongly plastic, and the front barely reaches $x \approx 260$ by $t = 30$. Here, we observe a near-stopping of the flow where the leading front propagates with an extremely low speed.

Our non-intrusive TROM reproduces the FOM simulations in all non-Newtonian regimes. In particular, the TROM model reproduces the speed of the propagating leading front and the plus/shear regions of the flow. Moreover, the simulations in this section demonstrate that the TROM model performs well in out-of-sample regimes, accurately capturing key properties of non-Newtonian flows.

%Agreement is tightest at later times when the reduced basis captures the dominant spatial modes; minor discrepancies in the earliest contour ($t = 5$) reflect the steep near-gate transients that demand a locally richer representation.

% ------------------------
%  FIGURES: τ_c = 300 Pa
% ------------------------

% ------------------------
\subsubsection{Effect of Bed $\hR$}
\label{subsubsec:hR_impact}
\begin{figure}[H]
  \centering
  \begin{subfigure}[b]{0.49\linewidth}
    \centering
    \includegraphics[width=\linewidth]{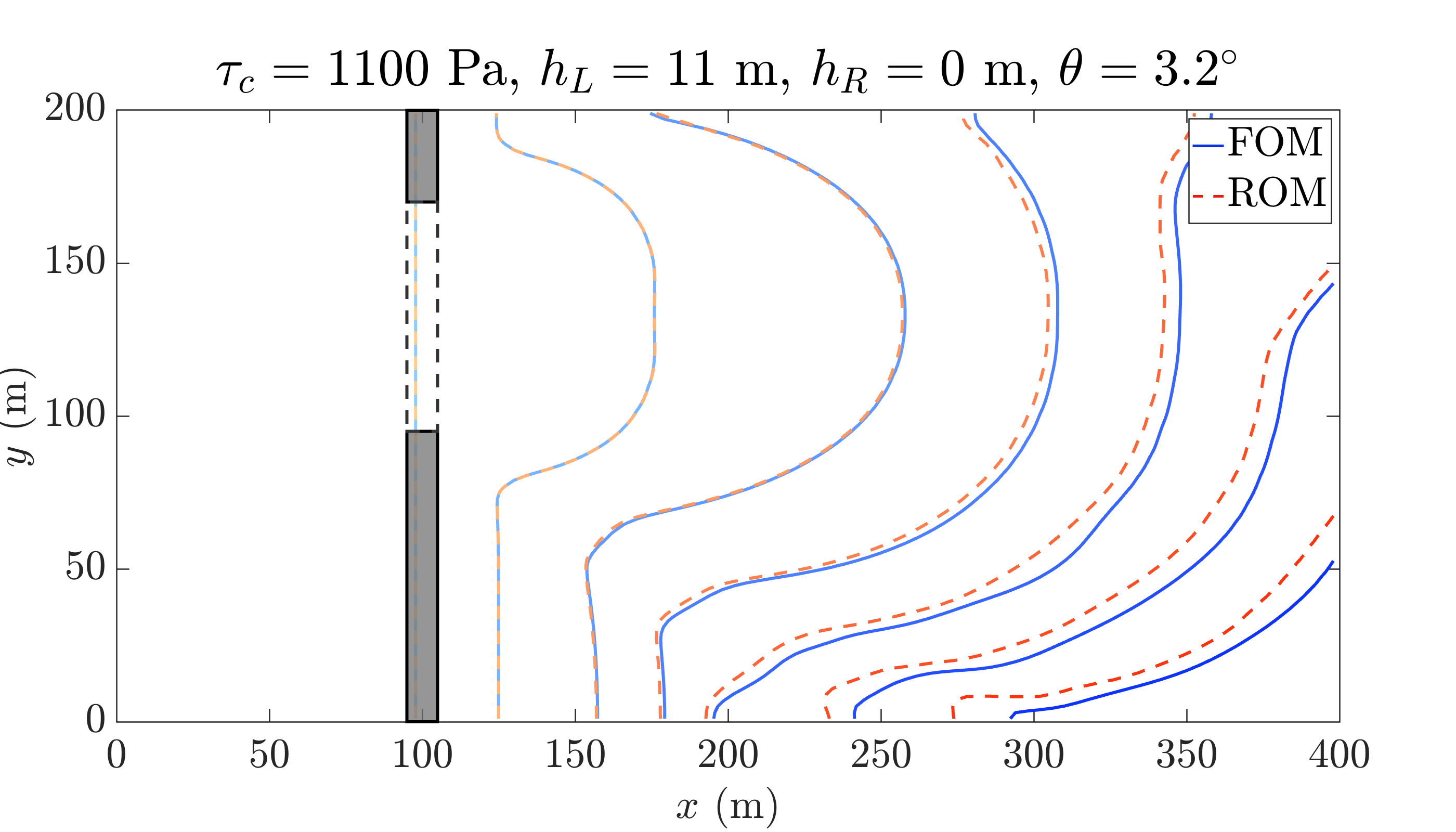}
    \caption{Flow front evolution, $t = 0, 5, \ldots, 30$.}
    \label{fig:front_tau1100_hR_0}
  \end{subfigure}
  \hfill
  \begin{subfigure}[b]{0.49\linewidth}
    \centering
    \includegraphics[width=\linewidth]{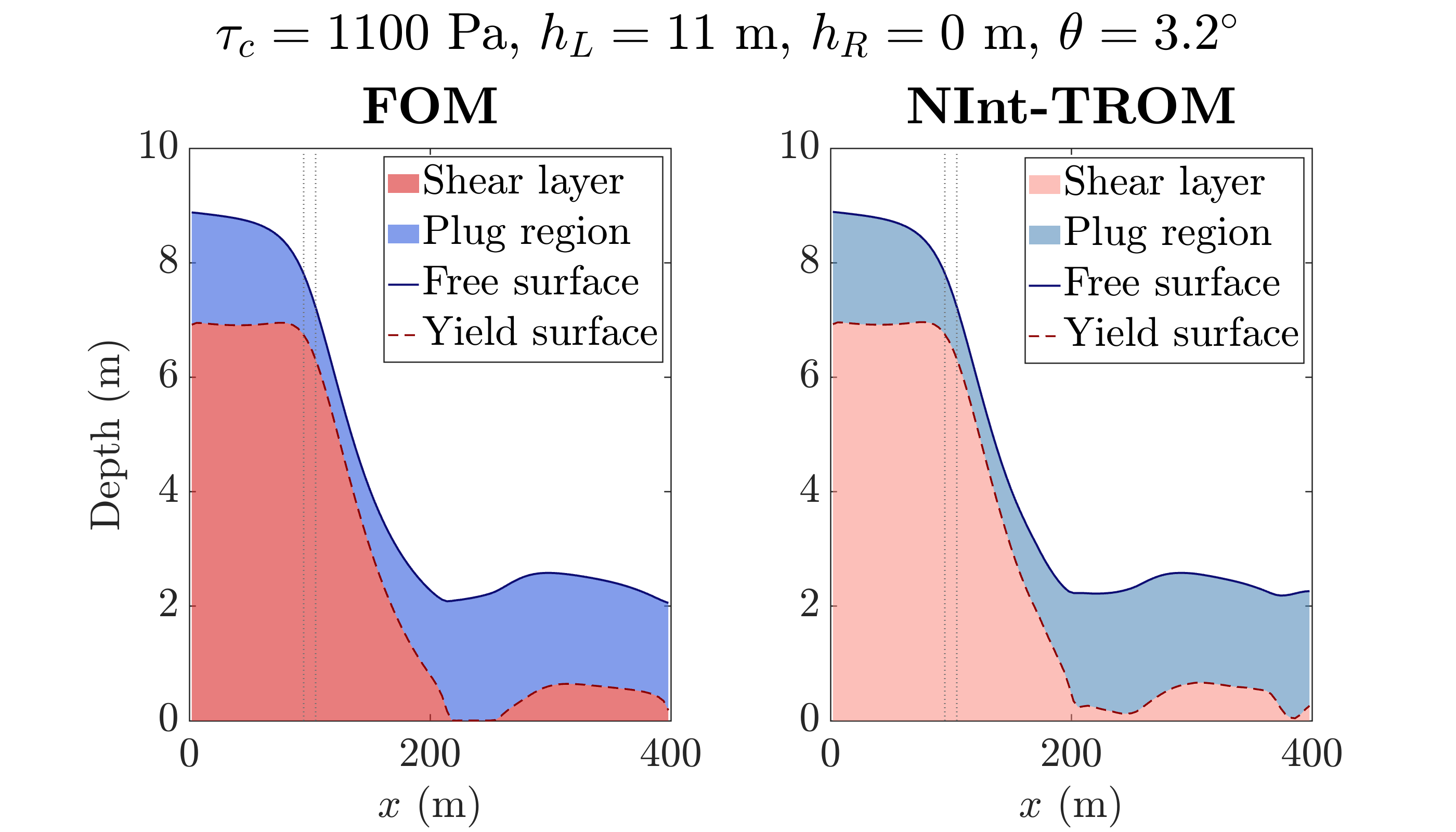}
    \caption{Flow front evolution, $t = 0, 5, \ldots, 30$.}
    \label{fig:xs_tau1100_hR_0}
  \end{subfigure}
  \caption{$\tauc = 1100$, $\hL = 11$, $\hR = 0.0$,
           $\thet = 3.2^{\circ}$.
           Blue solid: FOM; red dashed: NI-TROM.}
  \label{fig:tau1100_hR_0}
\end{figure}

\begin{figure}[H]
  \centering
  \begin{subfigure}[b]{0.49\linewidth}
    \centering
    \includegraphics[width=\linewidth]{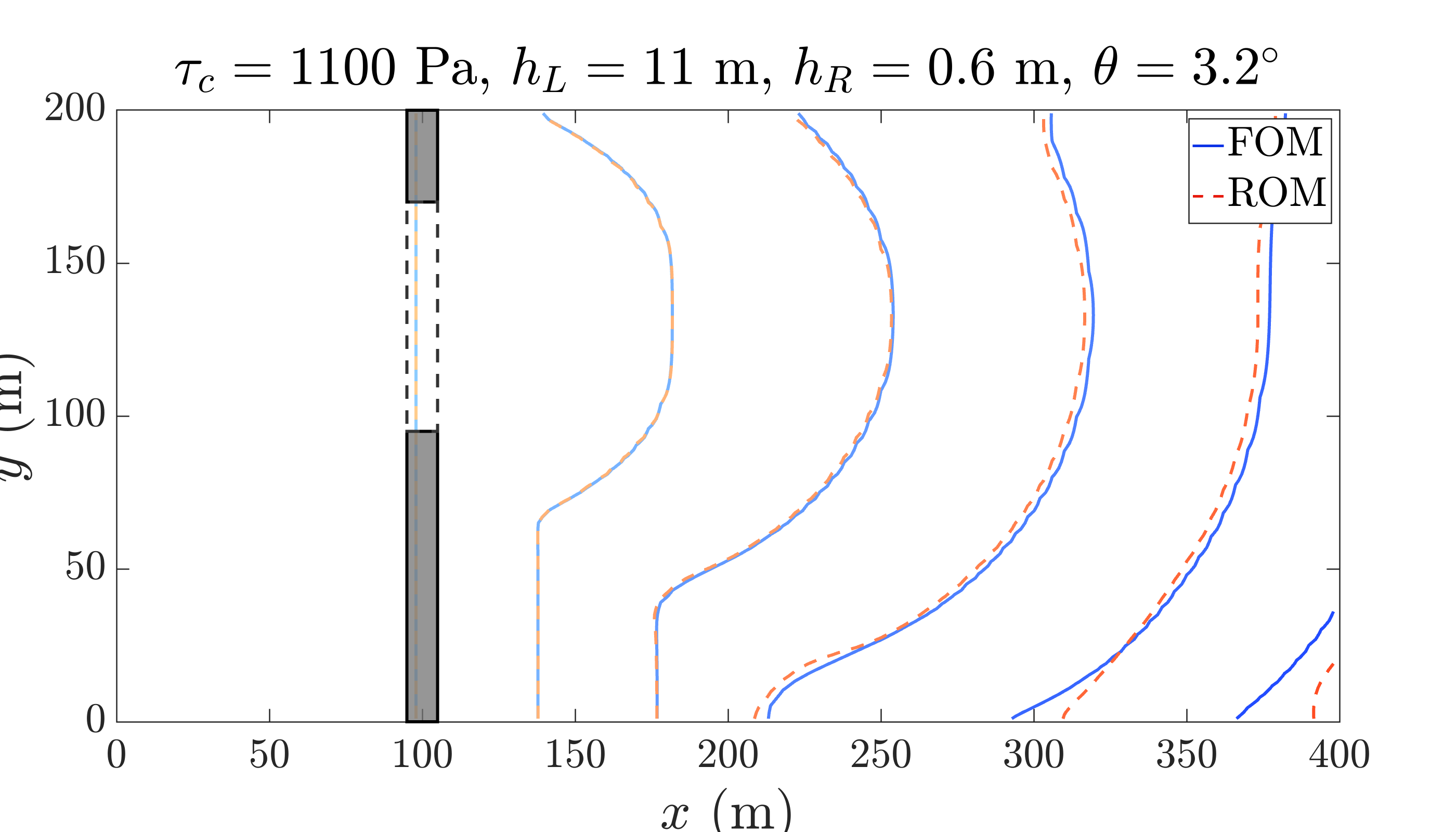}
    \caption{Flow front evolution, $t = 0, 5, \ldots, 30$.}
    \label{fig:front_tau1100_hR_0.6}
  \end{subfigure}
  \hfill
  \begin{subfigure}[b]{0.49\linewidth}
    \centering
    \includegraphics[width=\linewidth]{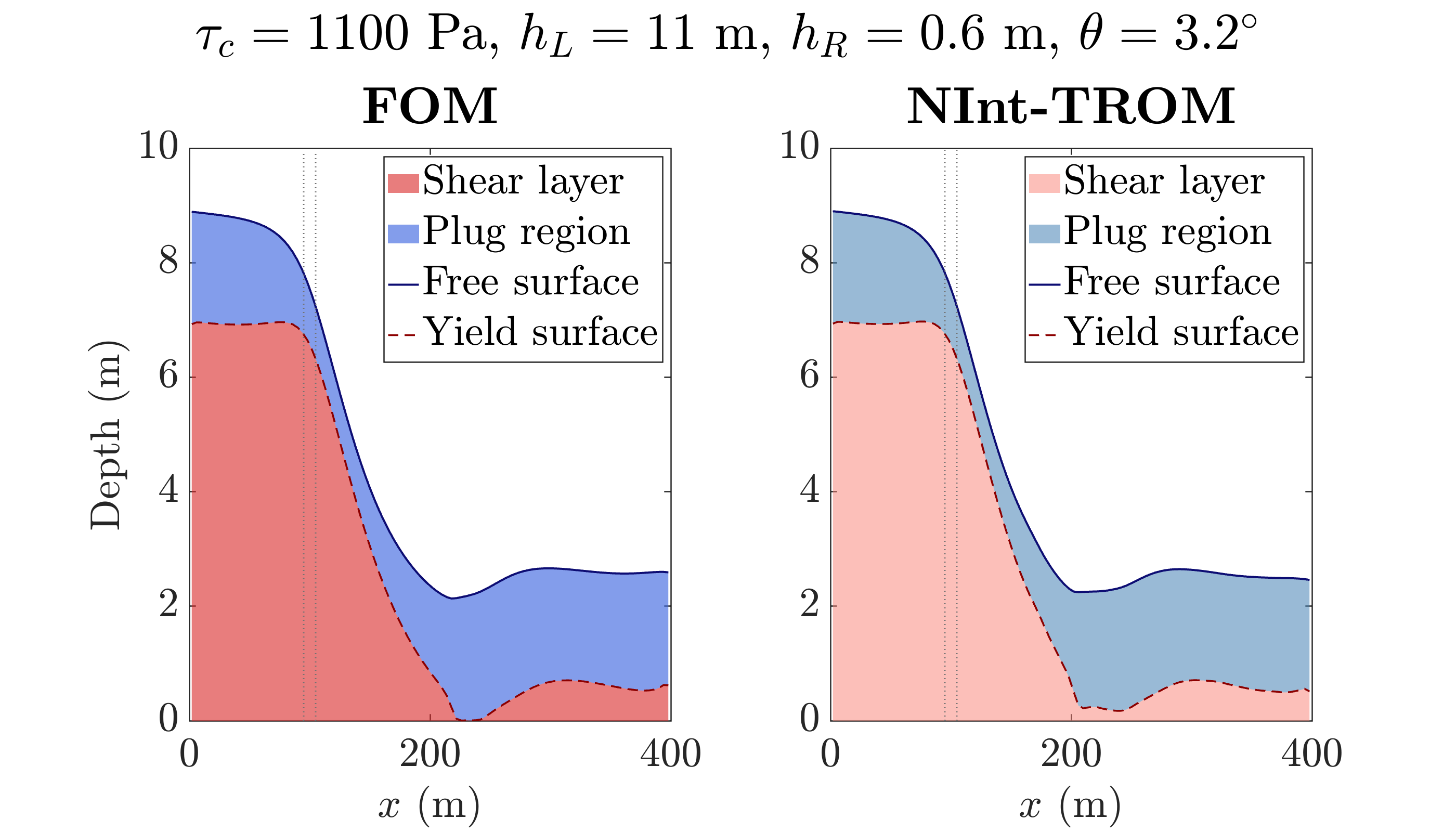}
    \caption{Flow front evolution, $t = 0, 5, \ldots, 30$.}
    \label{fig:xs_tau1100_hR_0.6}
  \end{subfigure}
  \caption{$\tauc = 1100$, $\hL = 11$, $\hR = 0.6$,
           $\thet = 3.2^{\circ}$.
           Blue solid: FOM; red dashed: NI-TROM.}
  \label{fig:tau1100_hR_0.6}
\end{figure}

\begin{figure}[H]
  \centering
  \begin{subfigure}[b]{0.49\linewidth}
    \centering
    \includegraphics[width=\linewidth]{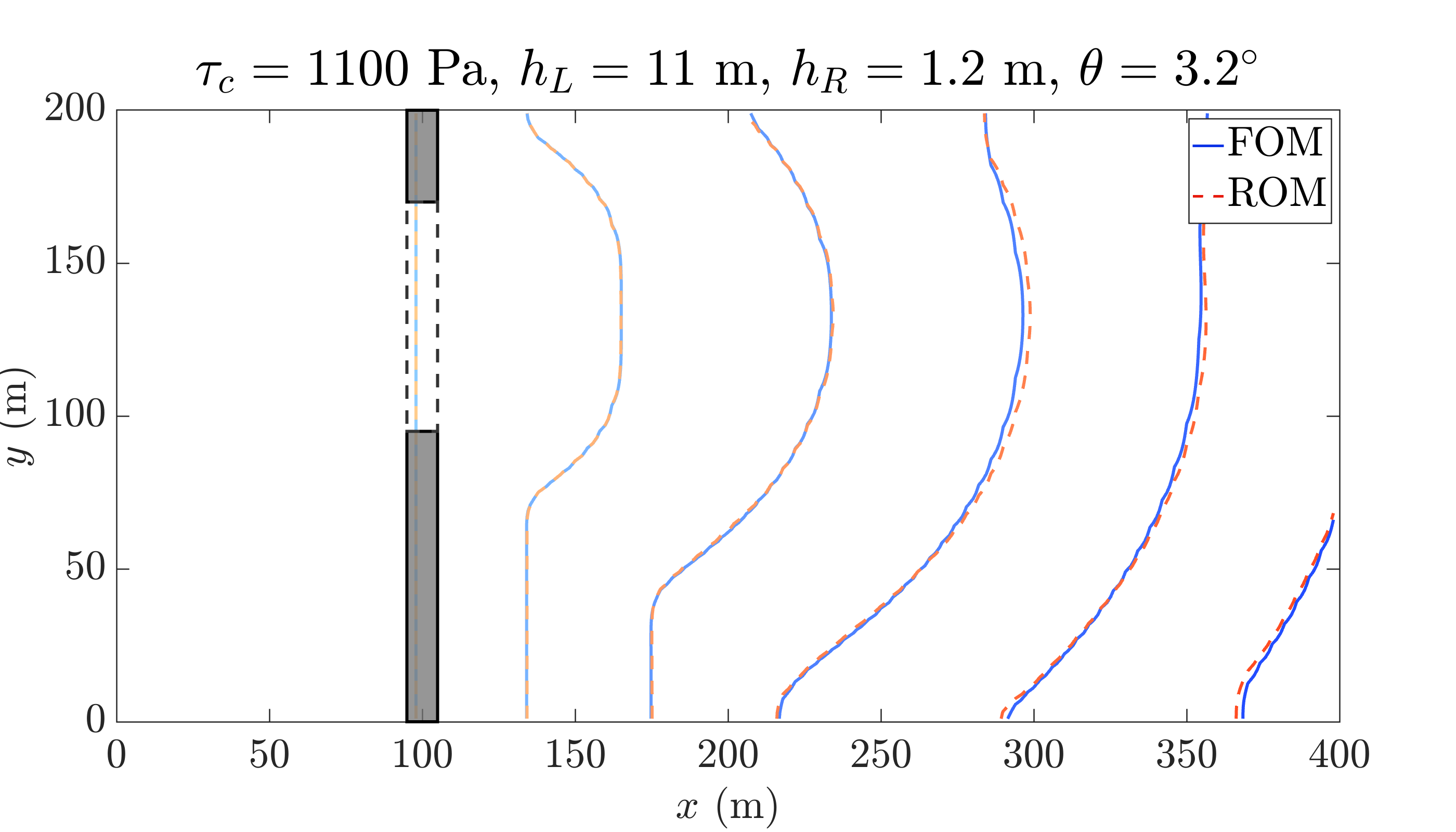}
    \caption{Flow front evolution, $t = 0, 5, \ldots, 30$.}
    \label{fig:front_tau1100_hR_1.2}
  \end{subfigure}
  \hfill
  \begin{subfigure}[b]{0.49\linewidth}
    \centering
    \includegraphics[width=\linewidth]{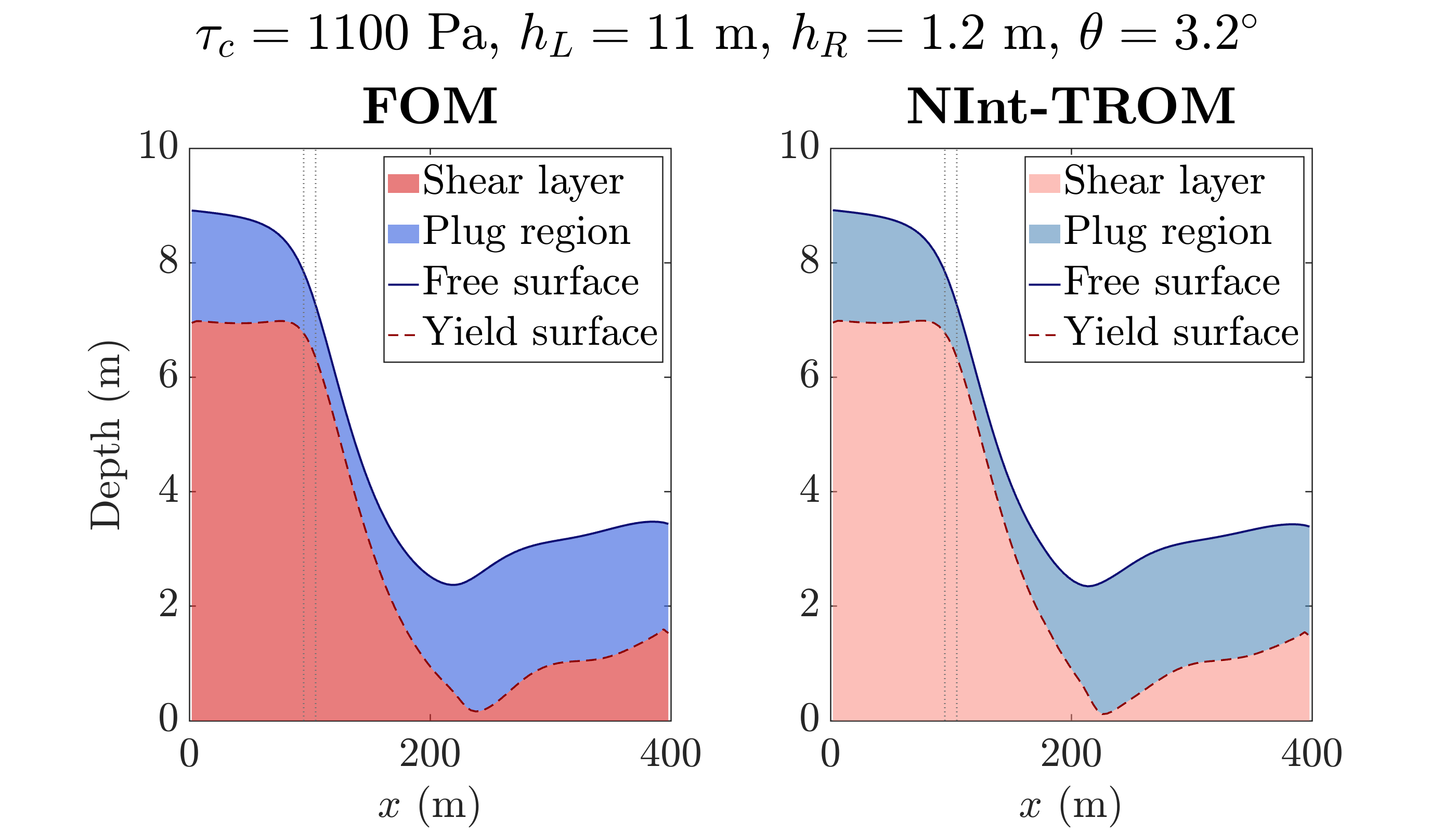}
    \caption{Flow front evolution, $t = 0, 5, \ldots, 30$.}
    \label{fig:xs_tau1100_hR_1.2}
  \end{subfigure}
  \caption{$\tauc = 1100$, $\hL = 11$, $\hR = 1.2$,
           $\thet = 3.2^{\circ}$.
           Blue solid: FOM; red dashed: NI-TROM.}
  \label{fig:tau1100_hR_1.2}
\end{figure}

% ------------------------
In this section, we analyze the performance of our TROM in a non-Newtonian regime with varying initial condition $\hR$. In particular, we fix parameters 
$\tauc=1100$, $\hL=11$, $\thet=3.2$ and vary 
$\hR=0, 0.6, 1.2$. Case $\hR = 1.2$ is identical to one considered in the previous section, but is repeated here to simplify presentation.
Figures \ref{fig:tau1100_hR_0}--\ref{fig:tau1100_hR_1.2} depict leading front contours and plug/shear regions for the three values of $\hR$. 
Here, we observe the importance of non-Newtonian effects, as the propagating front for $\hR=0$ is significantly slower than that for $\hR=0.6$ and $1.2$.
This is, most likely, due to the presence of a thicker shear region for $\hR=0.6$ and $1.2$ at the bottom of the propagating fluid. As the thickness of the shear region increases, it is "easier" for fluid to move along the inclined plane.
In addition, 
similar to the simulations of the Newtonian fluid described in Section \ref{subsubsec:newtonian}, the leading propagating front slows down slightly for $\hR=1.2$ compared to $\hR=0.6$.

Figures \ref{fig:tau1100_hR_0}--\ref{fig:tau1100_hR_1.2} demonstrate that our TROM reproduces both, the leading propagating wave and plug/shear regions well in all regimes. There are small discrepancies for $\hR=0$ that are more pronounced near the boundary $y=0$ during later times. This demonstrates that in this regime, TROM may be missing some small-scale features necessary to reproduce the multiscale nature of the solution near the boundary.
Regimes exhibiting fine multiscale features are more challenging, and discrepancies between the FOM and ROM are therefore expected, given that the ROM considered here is projection-based.
We would like to point out that TROM reproduced the solution with a similar multiscale structure near the boundary in the Newtonian regime very well  (Figs.~\ref{fig:front_dry_11} and~\ref{fig:front_dry_13}). 
Overall, 
TROM performs very well in the non-Newtonian regime considered in this section. In particular, TROM captures the most important features of the flow, such as the speed of the propagating wave and plug/shear regions.

% ------------------------
\subsubsection{Effect of Bed Slope $\theta$}
\label{subsubsec:theta_impact}
% ------------------------
To examine the influence of bed slope, we fix
$\hL = 11$, $\hR = 1.2$ and vary the slope angle
$\thet =0,\, 1.5,\, 3.2$
for two representative yield stresses, $\tauc = 3000$ and
$\tauc = 6000$.  The gravitational driving force along the
bed is proportional to $\sin\thet$, so increasing $\thet$
directly augments the downstream momentum source in the
governing equations~\eqref{eq:2d_hb_conservative}.

\begin{figure}[H]
  \centering
  \begin{subfigure}[t]{0.49\linewidth}
    \centering
    \includegraphics[width=\linewidth]
      {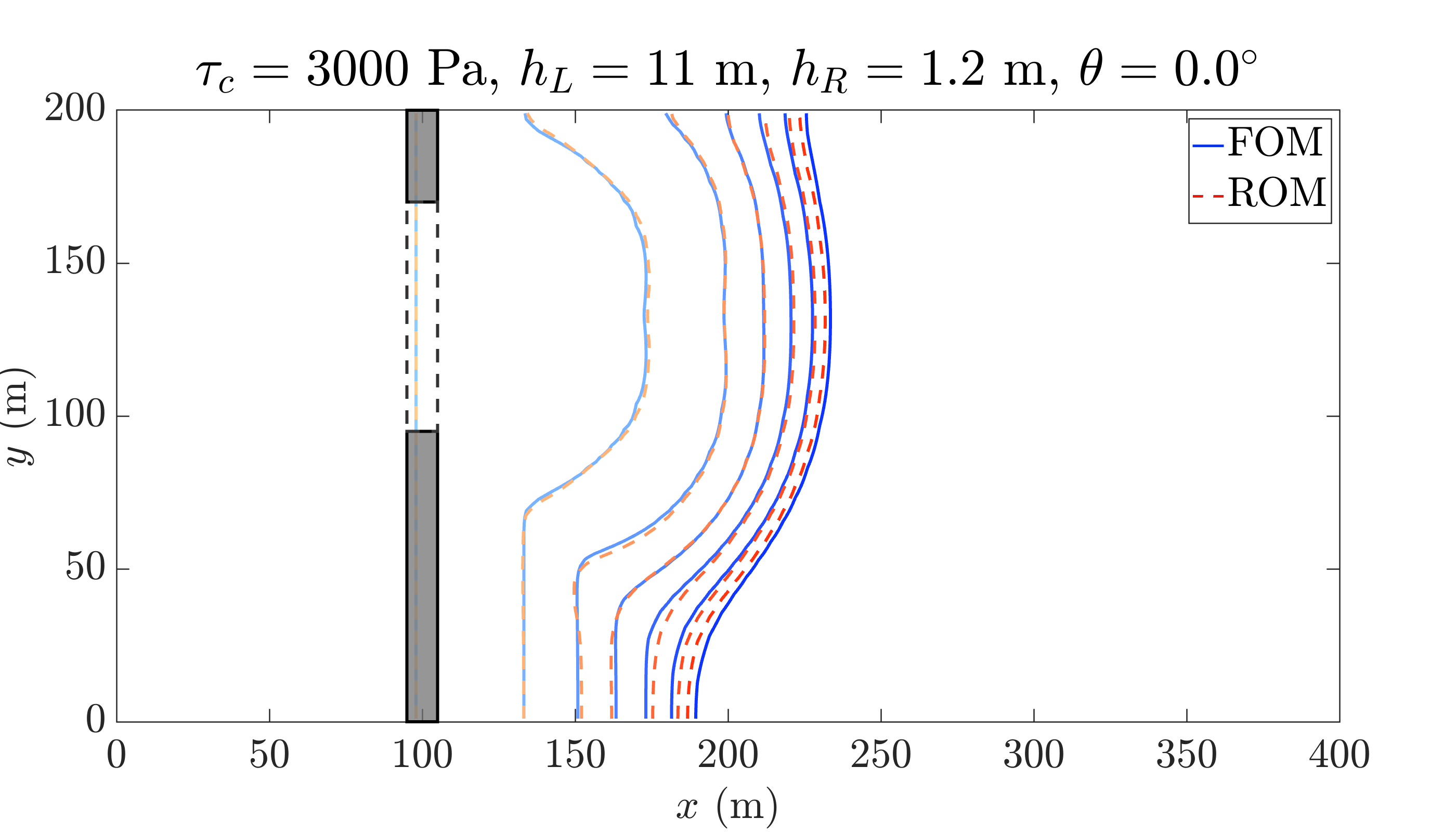}
    \caption{$\thet = 0.0$, snapshots at
             $t = 0, 5, \ldots, 30$.}
    \label{fig:front_tau3000_theta00}
  \end{subfigure}
  \hfill
  \begin{subfigure}[t]{0.49\linewidth}
    \centering
    \includegraphics[width=\linewidth]
      {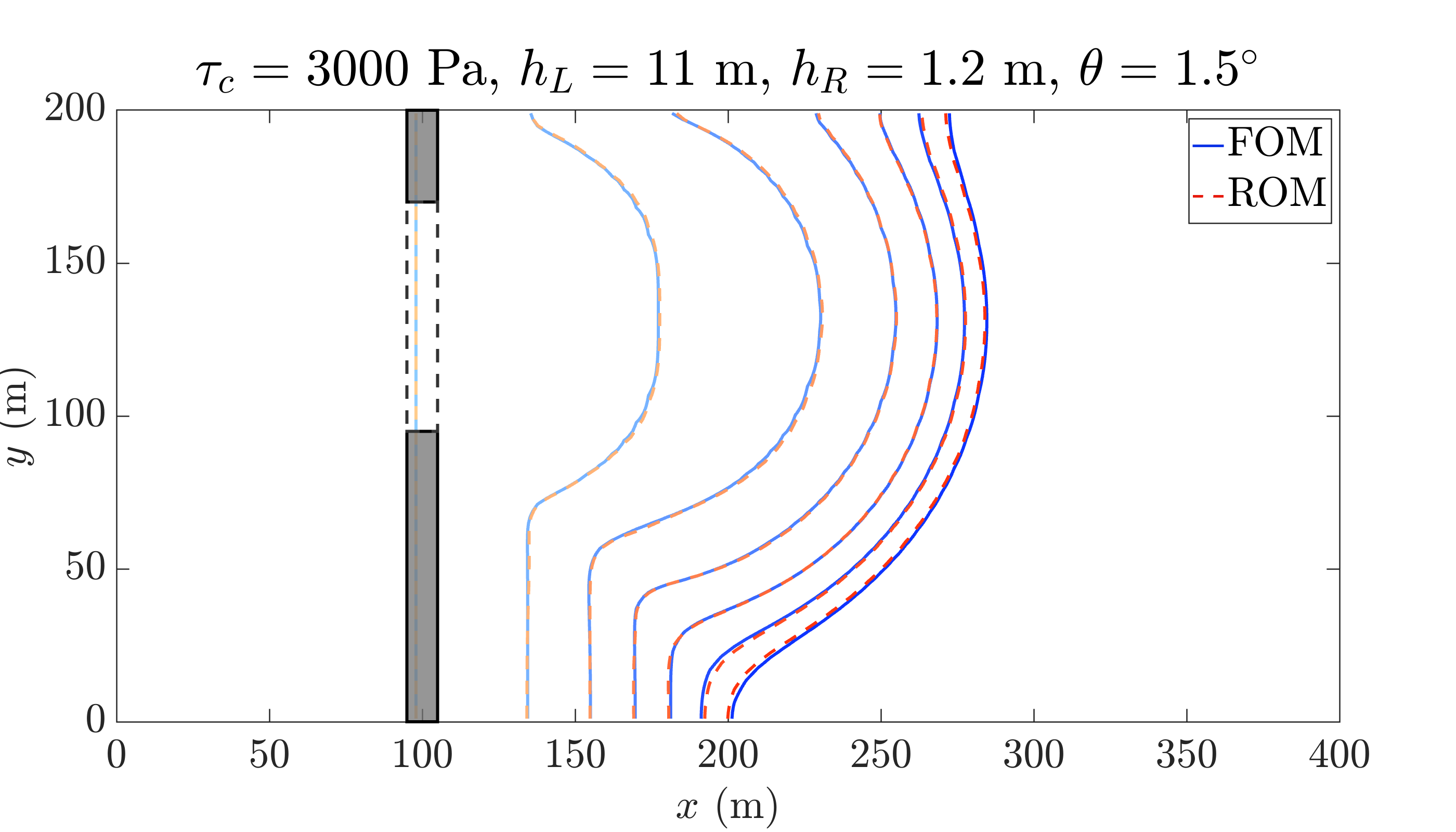}
    \caption{$\thet = 1.5$, snapshots at
             $t = 0, 5, \ldots, 30$.}
    \label{fig:front_tau3000_theta15}
  \end{subfigure}
  \caption{Effect of bed slope on flow front evolution:
           $\tauc = 3000$, $\hL = 11$, $\hR = 1.2$.
           The $\thet = 3.2$ case is shown in
           Fig.~\ref{fig:front_tau3000}.
           Blue solid: FOM; red dashed: NI-TROM.}
  \label{fig:front_theta_tau3000}
\end{figure}

\begin{figure}[H]
  \centering
  \begin{subfigure}[t]{0.49\linewidth}
    \centering
    \includegraphics[width=\linewidth]
      {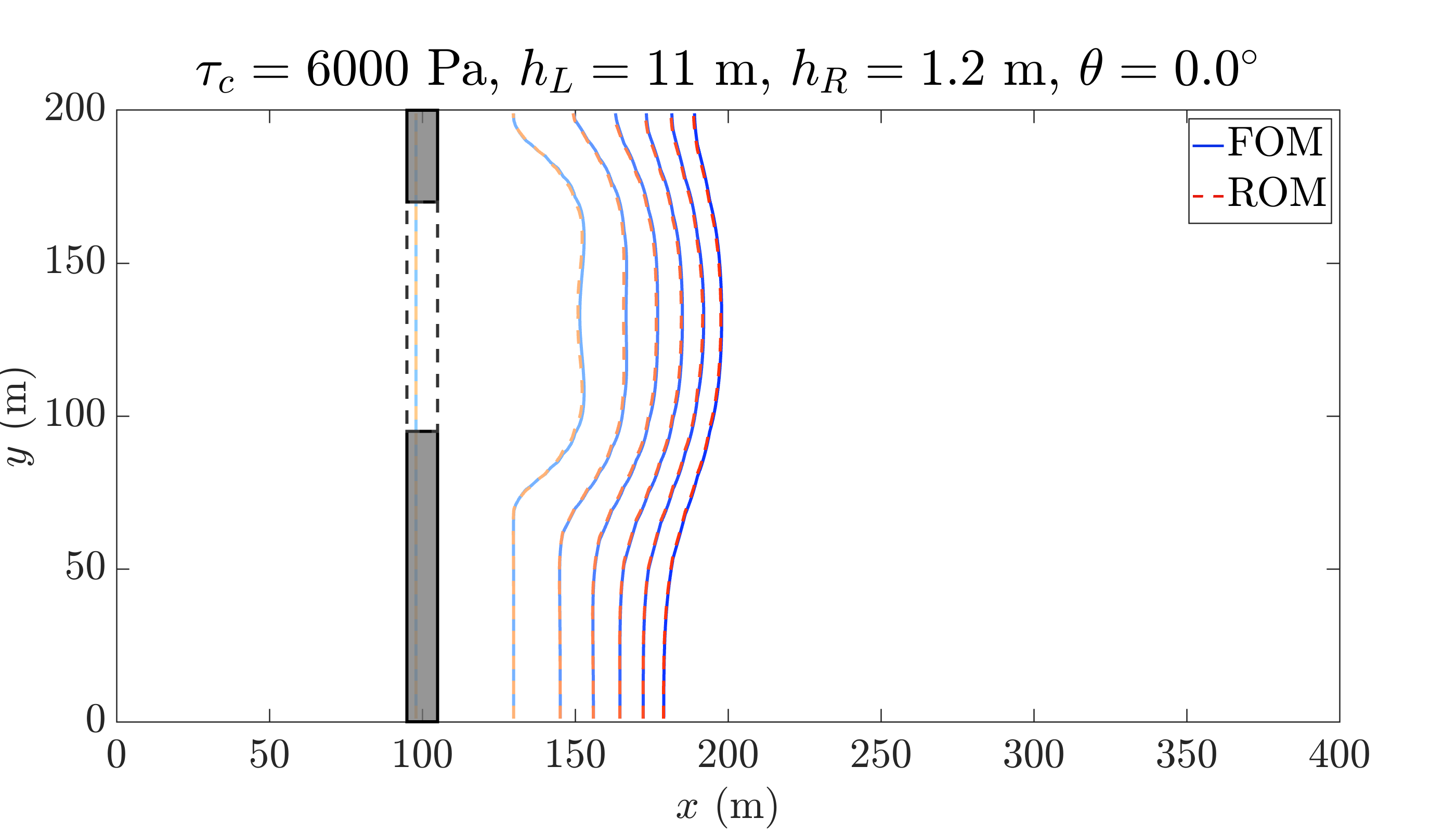}
    \caption{$\thet = 0.0^{\circ}$, snapshots at
             $t = 0, 5, \ldots, 30$.}
    \label{fig:front_tau6000_theta00}
  \end{subfigure}
  \hfill
  \begin{subfigure}[t]{0.49\linewidth}
    \centering
    \includegraphics[width=\linewidth]
      {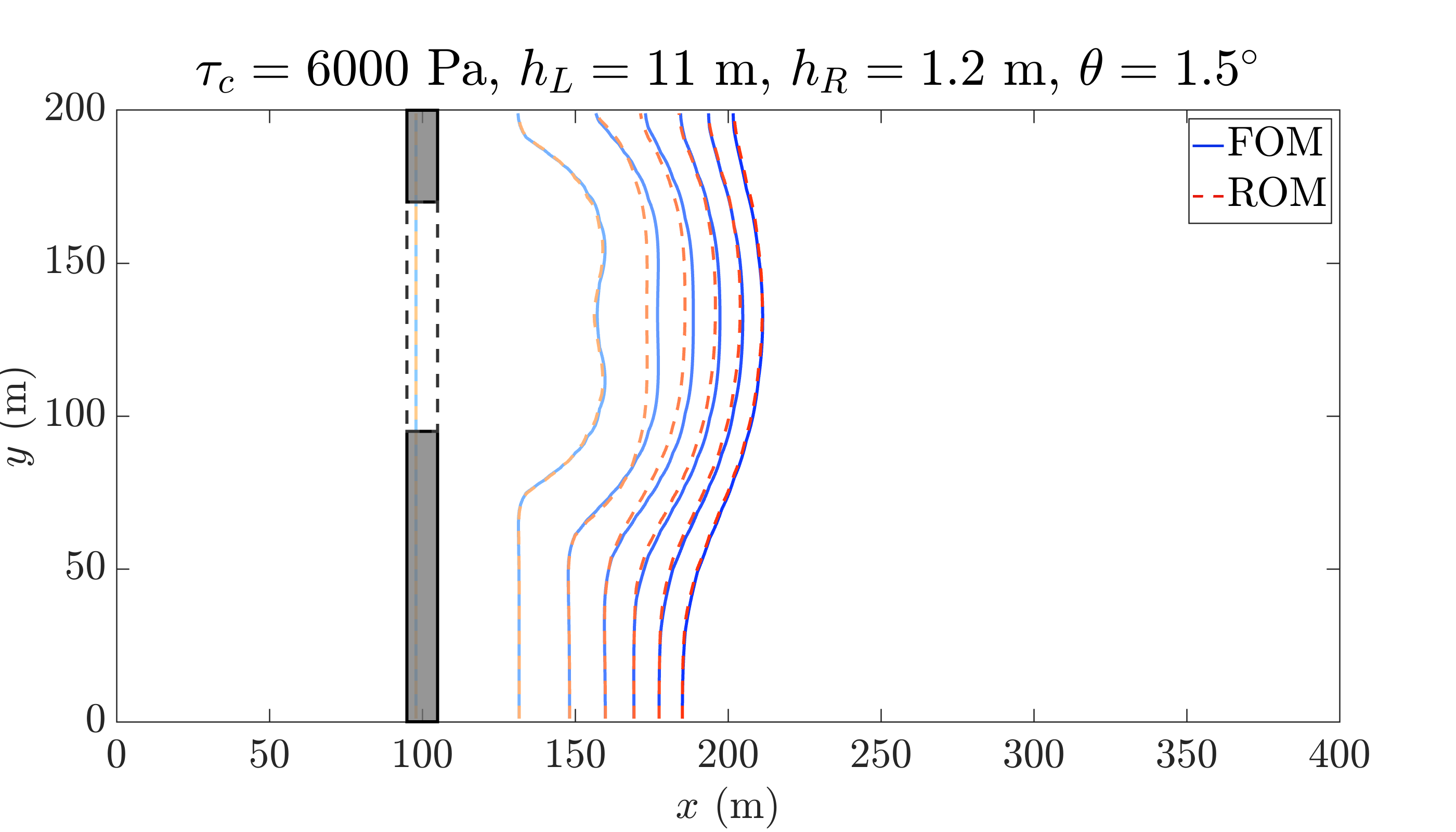}
    \caption{$\thet = 1.5^{\circ}$, snapshots at
             $t = 0, 5, \ldots, 30$.}
    \label{fig:front_tau6000_theta15}
  \end{subfigure}
  \caption{Effect of bed slope on flow front evolution:
           $\tauc = 6000$, $\hL = 11$, $\hR = 1.2$.
           The $\thet = 3.2^{\circ}$ case is shown in
           Fig.~\ref{fig:front_tau6000}.
           Blue solid: FOM; red dashed: NI-TROM.}
  \label{fig:front_theta_tau6000}
\end{figure}

Flow front contours for $\thet=0, 1.5$ are computed at
$t = 0, 5, 10, \ldots, 25, 30$ and presented in Figures \ref{fig:front_theta_tau3000} and \ref{fig:front_theta_tau6000} for 
$\tauc = 3000$ and
$\tauc = 6000$, respectively.
Results for $\thet=3.2$ are presented in Figures \ref{fig:front_tau3000} and \ref{fig:front_tau6000}.
Our numerical results demonstrate a monotone acceleration of the propagating wave with increasing slope.

The bed slope angle has a much stronger effect for $\tauc=3000$ than for the larger value $\tauc=6000$. In particular, comparing the simulations for the three slopes for $\tauc=3000$, the transition
from $\thet = 0$ to $\thet = 3.2$ more than doubles the
distance traveled by $t = 30$, emphasizing the dominant role of the bed geometry in determining the extent of the propagating wave.
For simulations with $\tauc=6000$, the differences 
between $\thet=0$, $\thet=1.5$, and $\thet=3.2$ are relatively small compared to those in simulations with $\tauc=3000$. For $\tauc=6000$, the propagating wave travels to approximately $x\approx 250$ for $\thet=3.2$, exhibiting a significant slowdown due to visco--plastic effects.
The increase in the leading front velocity due to the bed slope is therefore much weaker at high yield stress,
reflecting the dominance of plastic resistance over the
gravitational driving force. This competition between yield stress and bed slope is directly relevant to the
stopping-distance problem in geophysical flow hazards.

% ------------------------
%  FIGURES: τ_c = 3000 Pa, θ = 0.0° and θ = 1.5°
%  (θ = 3.2° already shown as Fig.~\ref{fig:front_tau3000})
% ------------------------

% -------------------------------------------------------
%  FIGURES: τ_c = 6000 Pa, θ = 0.0° and θ = 1.5°
%  (θ = 3.2° already shown as Fig.~\ref{fig:front_tau6000})
% -------------------------------------------------------

%%Speed Up

\subsection*{Computational Speedup}

The observed online speedup ranges from $87\times$ to $104\times$
across the test cases considered. More significant online time reduction is expected if  low dimensional statistics are of interest as the ROM output rather than the solution defined on a fine grid and for all times. Even greater reductions in online computational time can be expected when only low-dimensional statistics are of interest as ROM outputs, rather than the full solution defined on a fine grid over all time instances.

%=========================
\section{Conclusion}
\label{sec:conc}
In this paper, we present a non-intrusive Tensorial Reduced-Order Model applied to parametrized visco-plastic shallow free-surface flows. The main advantage of the non-intrusive TROM is that it does not require time integration of the underlying reduced equations, thus allowing for rapid evaluation of flow solutions at specified times. Moreover, the tensorial structure of the ROM naturally enables efficient parametric studies and many-query simulations of the underlying flow equations.

Our results demonstrate that the non-intrusive TROM performs very well for the dam-break problem for visco-plastic flows. In particular, the TROM accurately reproduces key flow features, including the propagation of the leading front and the plug/shear regions of the flow across all parameter regimes considered. The TROM also reproduces FOM simulations in the near-stopping regime, which is particularly important for practical CFD applications involving highly viscous or yield-stress fluids.

Our studies also highlight several interesting flow characteristics of visco-plastic fluids reported in Section \ref{subsubsec:hR_impact}. In particular, in moderate yield stress regimes (e.g., $\tauc = 1100$), non-zero $\hR$ (wet-bed setup) leads to acceleration of the fluid due to the presence of a larger shear region near the bottom boundary. A larger plug region for $\hR=0$ results in a significant slowdown of the leading front compared to simulations with $\hR>0$.

Simulations with varying $\thet$ for two choices of the yield stress parameter $\tauc=3000$ and $\tauc=6000$ in Section \ref{subsubsec:theta_impact} demonstrate a nonlinear interaction between bed inclination and visco-plastic effects. For smaller values of $\tau_c$, Newtonian-like effects are more dominant, and thus a larger bed angle leads to significant acceleration of the leading front. For larger $\tauc$, non-Newtonian effects become stronger, and the influence of $\thet$ diminishes.

Overall, TROM is a powerful framework for developing efficient reduced models for parametrized nonlinear flow problems. The non-intrusive version discussed in this paper is particularly suitable as a data-driven alternative to machine learning approaches, such as PINNs, while retaining a direct connection to high-fidelity CFD simulations. The proposed methodology enables fast online evaluation together with accurate representation of complex flow dynamics. Therefore, TROM has strong potential for use in many-query CFD applications, uncertainty quantification, optimization, and digital-twin frameworks for parametrized flow problems.

%=========================
\section*{Acknowledgments}  This work was partly supported by the National Science Foundation (NSF)
under the award  DMS-2309197 (RM \& MO). 

\appendix

\section{Long-time FOM Solution behavior}
\label{subsec:long_fom_solution}

\begin{figure}[H]
  \centering
  \begin{subfigure}[t]{0.49\linewidth}
    \centering
    \includegraphics[width=\linewidth]{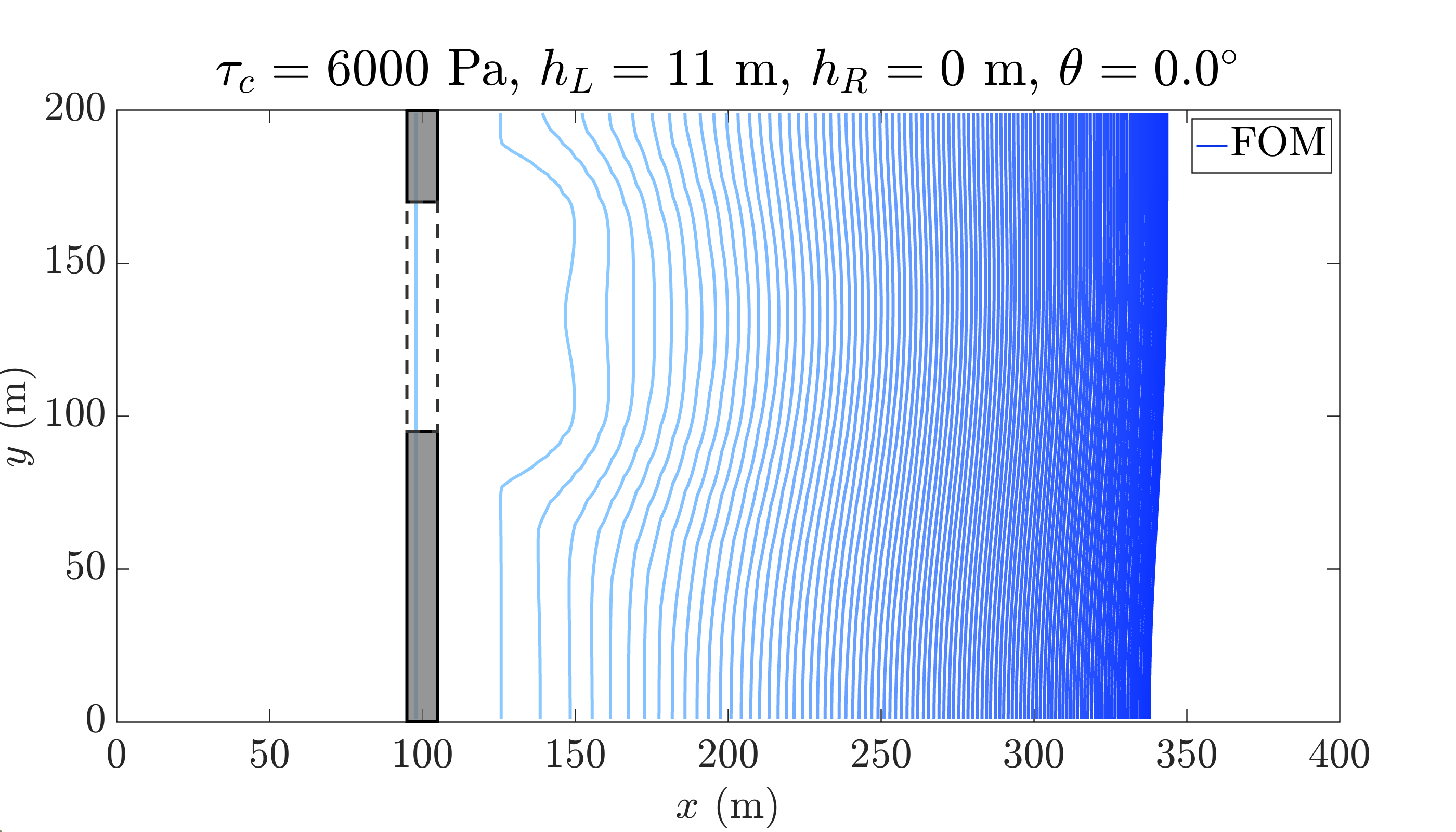}
    \caption{Dry bed ($\hR = 0$),
             $t =0, 5, 10, \ldots, 500$.}
    \label{fig:FOM_6000_0.0}
  \end{subfigure}
  \hfill
  \begin{subfigure}[t]{0.49\linewidth}
    \centering
    \includegraphics[width=\linewidth]{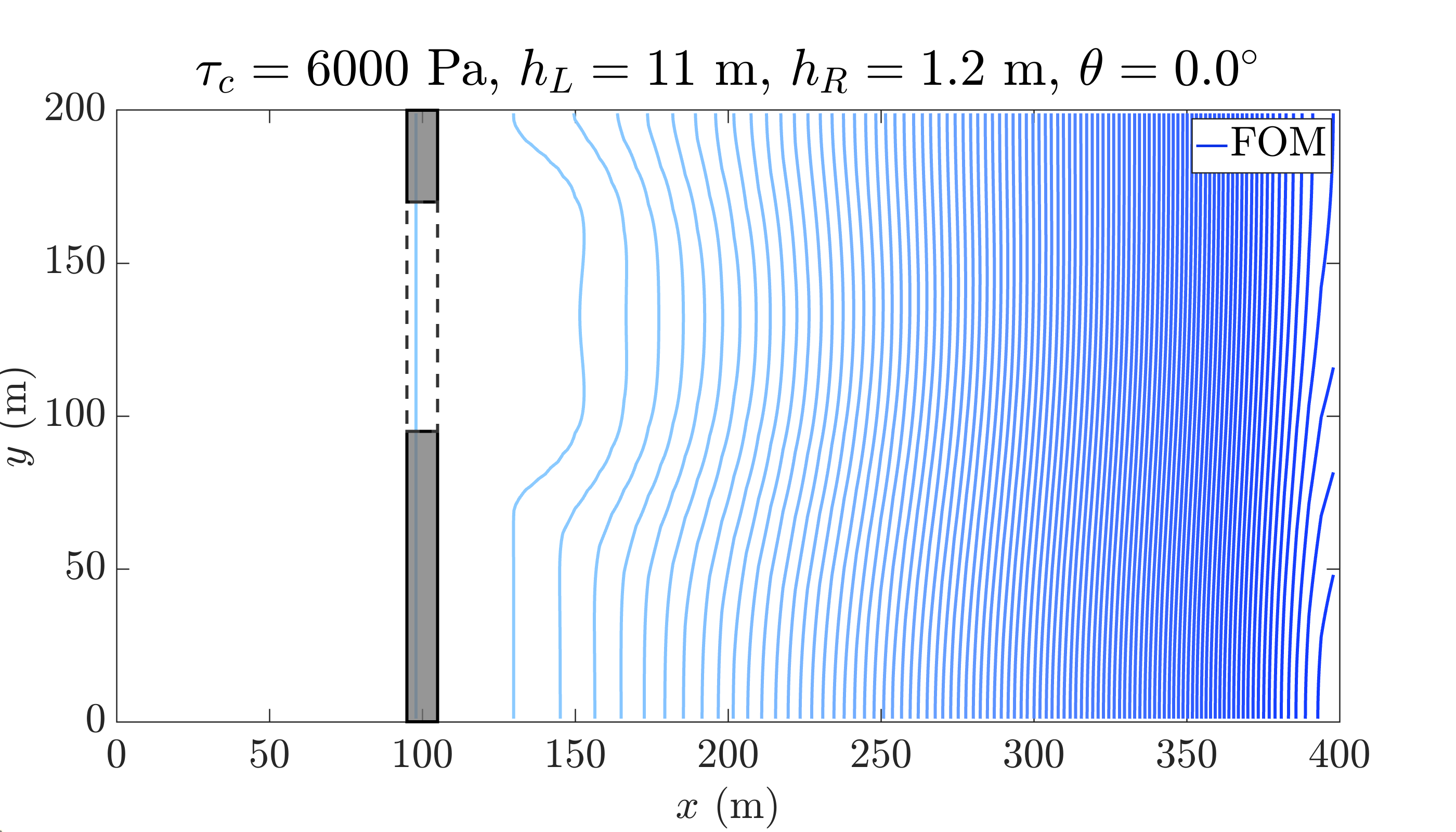}
    \caption{Wet bed ($\hR = 1.2$),
             $t = 0,5, 10, \ldots, 500$.}
    \label{fig:FOM_6000_1.2}
  \end{subfigure}
  \caption{Flow front evolution for the Long Run
           ($\tauc = 6000$, $\hL = 11$, $\thet = 0^{\circ}$).}
  \label{fig:front_FOM_6000}
\end{figure}

In this section, we briefly illustrate the long-time behavior of solutions in FOM simulations.
In particular, we consider the case $\tauc=6000$, $\thet=0$, $\hL=11$ and two values $\hR=0$ and $\hR=1.2$. The leading front propagation for the two values of $\hR$ is depicted in Figure \ref{fig:front_FOM_6000}. 
As discussed in this paper, the regime 
$\tauc=6000$ corresponds to strong non-Newtonian
effects. The flow exhibits a fast, inertia-dominated initial phase, then a sharp deceleration to a creeping-flow phase. For the dry-bed case $\hR=0$ (Figure \ref{fig:FOM_6000_0.0}), flow exhibits finite stopping time, while for the wet-bed simulations $\hR=1.2$ (Figure \ref{fig:FOM_6000_1.2}), the leading front keeps slowly propagating to the right and eventually reaches the boundary. Thus, wet-bed conditions have a non-trivial effect on visco--plastic flows. A recent study \cite{muchiri2024numerical}
also analyzed the dam-break problem for visco--plastic flows, and showed that for wet-bed initial conditions, the leading front propagates further downstream. 
Thus, we do not attempt to recover the final stopping time in this paper; instead, we focus on the TROM's ability to capture the fast--slow flow transition. 

% ========================
% REFERENCES
% ========================

%\bibliographystyle{plain}
%\bibliography{refs}

\end{document}